\documentclass[%
 reprint,
preprintnumbers,
 amsmath,amssymb,
 prd,
]{revtex4}

\newcommand{\lsim}   {\mathrel{\mathop{\kern 0pt \rlap
  {\raise.2ex\hbox{$<$}}}
  \lower.9ex\hbox{\kern-.190em $\sim$}}}
\newcommand{\gsim}   {\mathrel{\mathop{\kern 0pt \rlap
  {\raise.2ex\hbox{$>$}}}
  \lower.9ex\hbox{\kern-.190em $\sim$}}}

\usepackage{graphicx}% Include figure files
\usepackage{bm}% bold math
\usepackage[utf8]{inputenc}
\usepackage{slashed}
\usepackage{physics}
\usepackage[colorlinks=true,linktoc=all]{hyperref}% add hypertext capabilities
\usepackage{xcolor}
\usepackage{changebar}
\usepackage{mathtools}

\usepackage{mathrsfs}
\DeclareMathOperator{\diag}{diag}

\begin{document}

\title{TDiff invariant gauge fields in cosmology}% Force line breaks with \\
%\thanks{A footnote to the article title}

\author{Antonio L. Maroto}
\email{maroto@ucm.es}
\affiliation{Departamento de F\'{\i}sica Te\'orica and Instituto de F\'isica de Part\'iculas y del Cosmos (IPARCOS-UCM), Universidad Complutense de Madrid, 28040 Madrid, Spain}

\author{Alfredo D. Miravet}
 \email{alfrdelg@ucm.es}
\affiliation{Departamento de F\'{\i}sica Te\'orica and Instituto de F\'isica de Part\'iculas y del Cosmos (IPARCOS-UCM), Universidad Complutense de Madrid, 28040 Madrid, Spain}

\preprint{IPARCOS-UCM-24-014} % internal preprint number

%\collaboration{MUSO Collaboration}%\noaffiliation

\date{\today}% It is always \today, today,
             %  but any date may be explicitly specified

\begin{abstract}
We study the dynamics of Abelian gauge fields invariant under transverse diffeomorphisms (TDiff) in cosmological contexts.  We show that in the geometric optics approximation, very much as for Diff invariant theories,  the corresponding massless gauge bosons propagate along null geodesics and particle number is conserved. In addition, the polarization vectors are orthogonal to the propagation direction and the physical (transverse projection) polarization is parallel transported along the geodesics. We also consider TDiff invariant Dirac spinors, study the coupling to the gauge fields and analyze the conditions in order to avoid violations of Einstein's Equivalence Principle. The contributions to the energy-momentum tensor of the gauge field are also analyzed. We find that, in general, the breaking of Diff invariance makes the electric and magnetic parts of the vector field to gravitate in a different way. In the sub-Hubble regime we recover the standard radiation-like behaviour of the energy density, however in the super-Hubble regime the behaviour is totally different to the Diff case, thus opening up a wide range of possibilities for 
cosmological model building. In particular, possible effects on the evolution of large-scale primordial magnetic fields are discussed. 

\end{abstract}

%\keywords{Suggested keywords}%Use showkeys class option if keyword
                              %display desired
\maketitle

%\tableofcontents

\section{Introduction}

Einstein's General Relativity (GR) is the best description of gravity we have to date. It has performed exceptionally well in multiple tests ranging from the Solar System orbits to gravitational lensing and has been able to describe purely gravitational phenomena such as black holes and gravitational waves. It also serves as a theoretical framework for the standard Lambda-cold dark matter ($\Lambda$CDM) cosmology, which is a phenomenological model that accurately describes the large-scale structure and evolution of the universe with only a handful of parameters. GR relies on invariance under general coordinate transformations, i.e., invariance under diffeomorphisms (Diff) and  Einstein's Equivalence Principle \cite{Will:2014kxa}, which is equivalent to the Weak Equivalence Principle (WEP), Local Lorentz Invariance (LLI) and Local Position invariance (LPI). GR comes with its shortcomings though: Its lack of description of quantum gravity and the unknown fundamental nature of the dark sector of Cosmology have motivated the search for alternatives to GR. Generally, these modifications of GR consist of additional degrees of freedom, implemented in multiple ways, that alter the behaviour of gravity in a certain regime, typically at very long or very short distances. 

Over the last decade, there has been a growing interest in theories that break Diff-invariance down to invariance under transverse diffeomorphisms (TDiff), a subgroup restricted to volume-preserving general transformations. Early works include \cite{Alvarez:2006uu, Alvarez:2008zw}, in which a stability analysis of TDiff-invariant theories is performed at classical and one-loop levels. Here it was found that by enhancing the symmetry group with local Weyl invariance (dubbed WTDiff), the additional scalar degree of freedom is removed, thus propagating the same degrees of freedom as GR and preventing possible ghost instabilities. Unimodular gravity \cite{Einstein:1919gv, Unruh:1988in, Henneaux:1989zc, Kuchar:1991xd, Alvarez:2005iy, Jirousek:2018ago, Hammer:2020dqp, Carballo-Rubio:2022ofy}, the most popular TDiff-invariant theory for gravity, falls into this category and features the metric determinant treated as a non-dynamical scalar, so only the traceless part of Einstein's equations contribute to the dynamics. As a matter of fact, unimodular gravity has been proposed as a simple solution to the vacuum energy problem \cite{Ellis:2010uc}.  TDiff models beyond unimodular gravity have also been considered in  \cite{Alvarez:2006uu, Pirogov:2005im, Pirogov:2009hr, Pirogov:2011iq, Pirogov:2014lda}. In these models, the metric determinant is a dynamical field  and the corresponding spectrum includes a scalar graviton in addition to the standard massless spin-2 graviton. Also, the cosmological evolution in TDiff-invariant theories propagating a scalar graviton mode was recently investigated in \cite{Bello-Morales:2023btf}.

Breaking down to TDiff was also considered in the coupling to matter in the case of scalar fields in \cite{Alvarez:2007nn, Alvarez:2009ga} and possible violations of the Einstein Equivalence Principle were found. However, in \cite{Maroto:2023toq}, it was shown that, in the geometric optics approximation, when breaking down to TDiff invariance by a global factor in the matter action,   the three types of masses (inertial, active and passive) agree with those of standard Diff invariant theories thus avoiding the mentioned conflicts. In addition, it was found that in cosmological contexts on super-Hubble scales, these models exhibit a completely different behaviour of the energy-momentum tensor, making them very useful tools for the description of the dark sector. Thus in particular in \cite{Alonso-Lopez:2023hkx} a unified model of the dark sector with a single TDiff scalar field and the same number of free parameters as $\Lambda$CDM was presented.  TDiff-invariant scalar fields in arbitrary backgrounds have been considered in \cite{Jaramillo-Garrido:2023cor} and Diff symmetry restoration in \cite{Jaramillo-Garrido:2024tdv}. There it was shown that 
in the kinetic regime, these models behave as perfect adiabatic fluids and the corresponding speed of sound can be 
explicitly obtained. 

This work aims to study the dynamics of TDiff-invariant Abelian gauge fields in a cosmological context, following the novelties in scalar field dynamics found in \cite{Maroto:2023toq}. Vectors have been a topic in Cosmology since the pioneering works on vector-driven inflation \cite{Koivisto:2008xf, Maleknejad:2011jw, Maleknejad:2012fw}, and since then, they have been proposed on multiple occasions as candidates for the dark sector, chiefly for dark energy \cite{Armendariz-Picon:2004say, Boehmer:2007qa, BeltranJimenez:2008iye, BeltranJimenez:2008enx} or dark matter in the form of coherently-oscillating ultra-light vector fields \cite{Nelson:2011sf, LopezNacir:2018epg, Cembranos:2016ugq}.

The paper is organised as follows. In Sec. \ref{sec:general_dynamics}, we present our TDiff-invariant gauge field model and obtain the basic equations. In Sec. \ref{sec:geometric_optics}, we examine the geometric optics approximation for the vector field. In Sec. \ref{sec:external} we couple the field to an external current, and obtain the Lorentz force law by analysing the semiclassical limit of the coupling to a Dirac field. In Sec. \ref{sec:quantization}, we perform the canonical covariant quantization of the gauge field. In Sec. \ref{sec:homogeneous}, we turn our attention to an expanding universe and explore configurations of homogeneous electric and magnetic fields, as well as the conditions imposed by the conservation of the stress-energy tensor. In Sec. \ref{sec:general}, we consider a general electromagnetic field in an expanding universe and analyse the sub-Hubble regime. Finally, in Sec. \ref{sec:discussion}, we draw the main conclusions of the work.  We will work in natural units $\hbar=c=1$, with  metric signature $(+, -, -, -)$.

\section{General dynamics}\label{sec:general_dynamics}

Let us consider the most general action for an Abelian vector field $A_\mu(x)$ which is gauge invariant and invariant under transverse diffeomorphisms, to lowest order in field and metric derivatives and without parity violating terms,  \cite{Blas:2011ac}
\begin{equation}\label{eq:action}
    S_A =-\frac{1}{4}\int \dd[4]{x} f(g) F_{\mu\nu} F^{\mu\nu},
\end{equation}
where $F_{\mu\nu}=\partial_\mu A_\nu - \partial_\nu A_\mu$ is the field strength and $f(g)$ is an arbitrary positive function of the metric determinant $g=|\det(g_{\mu\nu})|$. The condition $f(g)>0$ ensures the action is free from ghosts or gradient instabilities \cite{Blas:2011ac,Rubakov:2014jja}. 

This action is invariant under a smaller group than full diffeomorphisms, which is the group of transverse diffeomorphisms (TDiff). In terms of infinitesimal coordinate transformations, these are implemented by
\begin{equation}
    x^\mu \to \tilde{x}^\mu = x^\mu + \xi^\mu,\quad \partial_\mu \xi^\mu = 0,
\end{equation}
so there is one less gauge degree of freedom. Restricting the symmetry group in this way allows us to write the action for any type of field (to the lowest order in metric derivatives)  with a general $f(g)$. Indeed, the variation of a general action 
\begin{equation}
    S = \int \dd[4]{x} f(g) \mathcal{L},
\end{equation}
under diffeomorphisms, with $\mathcal{L}$ a scalar function of the fields and their derivatives and the metric, can be shown to be \cite{Maroto:2023toq}
\begin{equation}
    \delta S = \int \dd[4]{x} \partial_\mu \xi^\mu [f(g)- 2gf'(g)] \mathcal{L},
\end{equation}
where the prime in $f'(g)$ denotes derivative with respect to its argument. This variation vanishes by either setting $f(g)=\sqrt{g}$, which grants invariance under the full group of diffeomorphisms, or by restricting ourselves to TDiff transformations $\partial_\mu \xi^\mu=0$. The Diff-invariant case can always be recovered by taking the limit $f(g)\to\sqrt{g}$. Notice that under TDiff transformations,   scalar densities (such as $g$) behave as pure scalars. 

Note that we break Diff invariance down to TDiff invariance only in this sector, while preserving full Diff-invariant actions in every other sector, in particular, in the Einstein-Hilbert action, so that the total action for the TDiff vector field coupled to gravity would be 
\begin{align}
S=S_{EH}+S_A=    -\frac{1}{16\pi G}\int d^4x \sqrt{g}\,R  -\frac{1}{4}\int \dd[4]{x} f(g) F_{\mu\nu} F^{\mu\nu}, \label{total_action}
\end{align}

We can obtain the equations of motion for the vector field by varying the action \eqref{total_action} with respect to it
\begin{equation}\label{eq:general_eom}
  \partial_\mu[f(g) F^{\mu\nu}] = 0. 
\end{equation}
Variations of the total action \eqref{total_action} with respect to the 
metric tensor yields the corresponding Einstein equation
\begin{align}
    R^{\mu\nu}-\frac{1}{2}g^{\mu\nu}R=8\pi G T^{\mu\nu}
\end{align}
where the stress-energy tensor, appearing on the right-hand side of the equation is obtained from the usual definition
\begin{align}
T^{\mu \nu}=-\frac{2}{\sqrt{g}}\frac{\delta {S_A}}{\delta g_{\mu\nu}}
\end{align}
For the action \eqref{eq:action}, it is given by
\begin{equation}
    T^\alpha{}_\beta = \frac{f(g)}{\sqrt{g}}\left(\frac{1}{2} f_1 F_{\mu\nu} F^{\mu\nu} \delta^\alpha{}_\beta - F^{\alpha\mu} F_{\beta\mu} \right), \label{energy-momentum}
\end{equation}
where we define
\begin{equation}
    f_n = \left(\frac{\dd}{\dd{\log g}}\right)^n \log f(g),
\end{equation}
and in particular, up to second-order derivatives, which will be extensively used in this work,
\begin{align}
    f_1 &= g \frac{f'}{f},\\ 
    f_2 &= f_1 - f_1^2 + g^2\frac{f''}{f},
\end{align}
so that in the Diff case $f(g)=\sqrt{g}$ we have $f_1=1/2$ and $f_n=0$ for $n>1$.
Because of the breaking of Diff invariance, the stress-energy tensor defined in \eqref{energy-momentum} is not necessarily conserved under solutions of the field equations of motion, i.e. $\nabla_\mu T^{\mu\nu}\neq 0$ and does not reduce to the canonical one in flat spacetime (see \cite{Alvarez:2007nn, Alvarez:2009ga} for a discussion). However, since Diff invariance is not broken in the gravity sector, Bianchi identities $\nabla_\mu G^{\mu\nu}=0$ ensure $\nabla_\mu T^{\mu\nu}=0$, i.e. the energy-momentum is conserved on solutions of the Einstein equations. Speaking in practical terms, the conservation of the stress-energy tensor will impose constraints between the different functions in the metric tensor, which cannot be fixed \textit{a priori} with a coordinate transformation due to having one less gauge degree of freedom in TDiff models.

\section{Geometric optics approximation}\label{sec:geometric_optics}

 The breaking of diffeomorphism invariance could have important implications for the consistency of the theory  \cite{Alvarez:2009ga,Maroto:2023toq}. The presence of the space-time 
 dependent function $f(g)$ could violate the Equivalence Principle
 since it is not guaranteed that vector field particles follow the geodesics of the space-time geometry. In order to analyze this problem, and following the same approach as done in \cite{Alvarez:2009ga,Maroto:2023toq} for scalar fields,  we will consider the geometric optics approximation \cite{Misner:1973prb, Maroto:2023toq}. Geometric optics works as long as the typical variation length of the field, i.e. the wavelength $\lambda$, is much shorter than any other relevant length $L$, such as the length at which the amplitude of the field varies or the typical variation length of the space-time. 

In this approximation, a mode of the vector potential can be written as a product of an exponential, which is rapidly changing, and a slowly evolving complex amplitude that is expanded perturbatively in powers of $\lambda/L$
\begin{equation}
    A_\mu(x) = \Re\left[(V_\mu + W_\mu\epsilon + \dots )e^{i\theta(x)/\epsilon}\right],
\end{equation}
where $\theta(x)$ is a real function, $\epsilon$ is a dummy power-counting parameter and the wavevector is $k_\mu = \partial_\mu \theta$. Note that the amplitude functions $V_\mu$, $W_\mu$... are allowed to depend both on position and on $k_\mu$. Terms of order $\mathcal{O}(1/\epsilon^n)$, $n>0$ are valid in the geometric optics approximation, whereas terms of order $\mathcal{O}(1)$, $\mathcal{O}(\epsilon),$ etc. are said to belong to post-geometric optics. As a result, only the functions $V_\mu$ and $\theta$ have no post-geometric optics corrections. 

With all of this established, let us delve into the equations of motion. Firstly, the action \eqref{eq:action} is invariant under a gauge transformation of the field
\begin{equation}
    A_\mu(x) \to A_\mu'(x) = A_\mu(x) + \partial_\mu \phi(x),
\end{equation}
with $\phi(x)$ a smooth function, as it leaves the field strength tensor unchanged. This allows us to use the Lorentz gauge condition, which can be written as
\begin{equation}
    \nabla_\mu A^\mu = \left[\frac{i}{\epsilon}k_\mu (V^\mu + W^\mu \epsilon + \dots) + (V^\mu + W^\mu \epsilon + \dots)_{;\mu} \right] e^{i\theta/\epsilon} = 0.
\end{equation}

At leading order $\mathcal{O}(1/\epsilon)$, this yields
\begin{equation}\label{eq:gauge_LO}
    k_\mu V^\mu = 0,
\end{equation}
which means that the amplitude of the field (or the polarization vector) is perpendicular to the direction of propagation. The next order 
\begin{equation}\label{eq:gauge_NLO}
    V^\mu{}_{;\mu} + ik_\mu W^\mu = 0,
\end{equation}
already belongs to the post-geometric regime $\mathcal{O}(1)$, and in fact shows a deviation from this perpendicularity for the second term in the expansion of the amplitude.

The equation of motion for the vector field \eqref{eq:general_eom} at leading order $\mathcal{O}(1/\epsilon^2)$ yields
\begin{equation}\label{eq:eom_LO}
    k_\mu V^\mu k^\nu - k_\mu k^\mu V^\nu = 0,
\end{equation}
which after applying the gauge condition \eqref{eq:gauge_LO} gives us the null condition of the rays
\begin{equation}\label{eq:eom_LO2}
    k_\mu k^\mu = 0.
\end{equation}
From this equation, we can write
\begin{align}
  (k_\mu k^\mu)_{;\alpha} = 0,
\end{align}
and since $k_\mu$ is the gradient of a scalar, we can commute covariant derivatives $k_{\mu;\nu} = k_{\nu;\mu}$ to find
\begin{align}
  k^\mu k_{\mu;\alpha} =  k^\mu k_{\alpha;\mu}= 0. 
\end{align}
This is nothing but the geodesic equation. Thus, we confirm that very much as in standard Diff-invariant electromagnetism, in TDiff theories, massless gauge bosons propagate along the space-time geodesics in the geometric optics approximation. 

The next-to-leading order equation of motion gives us the following expression
\begin{equation}
    \partial_\mu\left(\frac{f}{\sqrt{g}}\right) (k^\mu V^\nu - k^\nu V^\mu) \\
   + \frac{f}{\sqrt{g}}\left[ (k^\mu V^\nu)_{;\mu} - (k^\nu V^\mu)_{;\mu} + k_\mu V^{\nu;\mu} - k_\mu V^{\mu;\nu} + i k_\mu k^\mu W^\nu - i k_\mu k^\nu W^\mu  \right]=0.
\end{equation}

After using the gauge condition \eqref{eq:gauge_LO} and the previous order equation \eqref{eq:eom_LO2}, this expression can be simplified to
\begin{equation}
    \frac{\sqrt{g}}{f} \partial_\mu\left(\frac{f}{\sqrt{g}}\right)  (k^\mu V^\nu - k^\nu V^\mu)  + V^\nu k^\mu{}_{;\mu} + 2k^\mu V^\nu{}_{;\mu} = 0, \label{amplit}
\end{equation}
where we have also used the commutation of  covariant derivatives $k_{\mu;\nu} = k_{\nu;\mu}$. 

By decomposing the amplitude into a (real) magnitude $V$ and a (complex)
polarization vector $v^\mu$, i.e. $V^\mu=Vv^\mu$ with $v^\mu v^*_\mu=-1$ and
$V^\mu V^*_\mu=-V^2$, this expression allows us to obtain the equation for the propagation of the magnitude and polarization. Thus, if we contract this equation with $V^*_\nu$ (and add the corresponding complex conjugate) we can further simplify the expression and find a conserved current
\begin{equation}
    \nabla_\mu\left(\frac{f}{\sqrt{g}} V^2 k^\mu \right)=0. \label{cons}
\end{equation}
Thus, very much as in the Diff invariant case, we can write an adiabatic invariant which can be identified with the conserved particle number. Notice however that the conserved current is modified by a $f(g)/\sqrt{g}$ factor which is a typical feature of TDiff-invariant theories, and it was already found in the scalar case in  \cite{Maroto:2023toq}. This is simply due to the fact that the action is changed by a factor $f/\sqrt{g}$, which also appears naturally in the derivation of Noether currents.

Finally, introducing the decomposition in magnitude and polarization in \eqref{amplit} and using \eqref{cons} we get the propagation equation for the polarization vector
\begin{align}
 k^\mu v^\nu_{\; ;\mu}=\frac{1}{2}\frac{\sqrt{g}}{f}\partial_\mu\left(\frac{f}{\sqrt{g}}\right)k^\nu v^\mu.
\end{align}
We see that, unlike the Diff-invariant case, in which the polarization vector is parallel transported along the geodesic $( k^\mu v^\nu_{\; ;\mu}=0)$, in the TDiff case,  the polarization vector changes. However, the change takes place along the (unphysical) longitudinal $k^\mu$ direction so that the conditions $k_\mu v^\mu=0$ and $v^\mu v^*_\mu=-1$ will be satisfied along the entire ray. In particular, for any vector field $l_\mu$ which is parallel-transported along the geodesic ($k^\mu l_{\nu;\mu}=0$) and is orthogonal to 
$k_\mu$, ($k_\mu l^\mu=0$), the projected polarization remains constant along the geodesic since
\begin{align}
    k^\mu(v^\nu l_\nu)_{;\mu}=(k^\mu v^\nu_{;\mu})l_\nu+k^\mu l_{\nu;\mu}v^\nu = \frac{1}{2}\frac{\sqrt{g}}{f}\partial_\mu\left(\frac{f}{\sqrt{g}}\right)k^\nu v^\mu l_\nu=0
\end{align}

In conclusion, in TDiff theories in the geometric optics approximation, massless vector bosons propagate along null geodesics, the polarization vectors are orthogonal to the propagation direction, particle number is conserved and the physical (transverse projection) polarization is parallel transported along the geodesic.

\section{Coupling to an external current}\label{sec:external}

So far we have considered the free gauge field, let us now consider a general coupling to an external source $j^\mu$
\begin{equation}
    S = -\int \dd[4]{x} \left( f(g) \frac{1}{4} F_{\mu\nu} F^{\mu\nu} + f_D(g) j^\mu A_\mu\right),
\end{equation}
with $f_D(g)$ an arbitrary function of the metric determinant and $j^\mu$ an external Diff vector current that we assume does not depend on the vector field $A_\mu$. Imposing gauge invariance of the full action requires the external current to satisfy the conservation equation given by 
\begin{equation}
    \nabla_\mu \left(\frac{f_D(g)}{\sqrt{g}} j^\mu\right) = 0.
\end{equation}
Thus we see that, as we have already seen in Section \ref{sec:geometric_optics}, conserved currents are modified by a factor $f_D(g)/\sqrt{g}$ when breaking Diff invariance down to TDiff invariance.

The equation of motion for the vector field is then modified to
\begin{equation}\label{eq:vf_coupled_eom}
    \nabla_\mu\left[ \frac{f(g)}{\sqrt{g}} F^{\mu\nu} \right] = \frac{f_D(g)}{\sqrt{g}} j^\nu.
\end{equation}
which can be rewritten as
\begin{align}
  \nabla_\mu F^{\mu\nu}+\frac{\sqrt{g}}{f}\partial_\mu\left(\frac{f}{\sqrt{g}}\right)F^{\mu\nu} = \frac{f_D(g)}{f(g)}j^\nu.   
\end{align}
Notice that in the adiabatic approximation in which terms involving metric derivatives are negligible compared to those involving derivatives of the field strength, the equations of motion reduce to 
\begin{align}
  \nabla_\mu F^{\mu\nu} \simeq \frac{f_D(g)}{f(g)}j^\nu,
\end{align}
which agree with the standard Diff expression for the Maxwell equations but only if $f_D(g)=f(g)$. In general, $f_D(g)\neq f(g)$ could induce a space-time dependence of 
the fermionic charges in contradiction with Local Position Invariance (LPI). 
In the following, we will present 
the explicit construction of the conserved current from TDiff invariant Dirac spinors.

\subsection{TDiff Dirac spinors and the Lorentz force}\label{subsec:dirac}

Let us now consider the particular case of a Dirac field $\Psi$ that couples to the gauge field, which will allow us to obtain the Lorentz force law by analysing its semiclassical limit. The Dirac action can be written in an explicitly self-adjoint way \cite{Birrell:1982ix} as
\begin{equation}\label{eq:dirac_action}
    S_D[\Psi] = \int \dd[4]{x} f_D(g) \left[\frac{i}{2}(\bar\Psi \gamma^\mu D_\mu \Psi - D_\mu \bar\Psi \gamma^\mu \Psi) - m\bar\Psi\Psi \right],
\end{equation}
where $\bar\Psi = \Psi^\dagger \gamma^0$ (this $\gamma^0$ is the usual gamma matrix used in flat spacetime) is the Dirac adjoint and the covariant derivatives are
\begin{align}
    D_\mu \Psi &= \mathfrak{D}_\mu \Psi + iq A_\mu \Psi,\\
    D_\mu \bar\Psi &= \mathfrak{D}_\mu \bar\Psi - iq A_\mu \Psi,
\end{align}
and we are not worrying about the details of the gamma matrices $\gamma^\mu$ or the covariant derivative $\mathfrak{D}_\mu$ yet. 
%The change of sign in the gauge connection ensures that $\bar\Psi\Psi$ behaves as a scalar. 
We can easily identify the current that couples to the electromagnetic field as
\begin{equation}
    j^\mu = q\bar\Psi \gamma^\mu \Psi.
\end{equation}

As shown before if the gauge  function $f(g)$ agrees with 
the fermion function $f_D(g) = f(g)$, the Dirac current sources the gauge field exactly as in the Diff-invariant case, so we make this choice from now onwards.

Before obtaining the TDiff version of Dirac's equation, we need to take care of the details related to how fermions behave in curved spacetime. It is convenient to introduce the \emph{vierbein} $e_a^\mu$ \cite{Nakahara:2003nw}, which allows us to use a non-coordinate basis $\{ \hat{e}_a\} = \{ e_a^\mu \partial_\mu \}$ defined so that it is orthogonal with respect to the spacetime metric $g_{\mu\nu}$, i.e.
\begin{equation}
    e^\mu_a e^\nu_b g_{\mu\nu} = \eta_{ab}, \qquad g_{\mu\nu} = e_\mu^a e_\nu^b \eta_{ab},
\end{equation}
with $\eta_{ab} = \diag(+,-,-,-)$ the flat metric and $e_\mu^a$ the inverse of the vierbein (with respect to both types of indices)
\begin{equation}
    e^\mu_a e^a_\nu = \delta^\mu_\nu, \qquad e^\mu_a e^b_\mu = \delta^b_a.
\end{equation}
where we use Latin indices $a,b$ for local Lorentz tensors and Greek
indices $\mu,\nu, \dots$ for generally covariant tensors. 
Let $\{\gamma^a\}$ be the Dirac matrices in flat spacetime, which satisfy the anticommutation relations $\{\gamma^a, \gamma^b\} = 2\eta^{ab}$. We can generalise these matrices to curved spacetime by defining $\gamma^\mu = e^\mu_a \gamma^a$, which can be found to satisfy
\begin{equation}
    \{ \gamma^\mu, \gamma^\nu \} = 2 g^{\mu\nu}.
\end{equation}

The full covariant derivative of a spinor field, including the gauge connection term, can be written as 
\begin{equation}
    D_\mu\Psi = \partial_\mu \Psi + \Gamma_\mu \Psi + iqA_\mu \Psi,
\end{equation}
where $\Gamma_\mu$ is the spin connection \cite{Nakahara:2003nw, Shapiro:2016pfm, Dobado:1997jx}, which implements the covariant derivative $\mathfrak{D}$ for objects which are defined on the Lorentz frame (those with flat indices)
\begin{equation}
    \mathfrak{D}_\mu \Psi = \partial_\mu \Psi + \Gamma_\mu \Psi.
\end{equation}
The specific form of the spin connection can be found by imposing the proper transformation laws under a local Lorentz transformation, which is
\begin{equation}
    \Gamma_\mu = -\frac{i}{2} \Gamma^a{}_\mu{}^b \Sigma_{ab},
\end{equation}
where $\Sigma_{ab} = \frac{i}{4}[\gamma_a, \gamma_b]$ is the spinor representation of the generators of the Lorentz transformations. We have also chosen the connection to be metric and torsion-free, so it is given by the Christoffel symbols. The connection with flat indices is defined by
\begin{equation}
    \nabla_a \hat{e}_b = \Gamma^c{}_{ab} \hat{e}_c,
\end{equation}
and it relates to the connection with curved indices as follows
\begin{equation}
    \Gamma^c{}_{ab} = e^c_\nu e_a^\mu \nabla_\mu e^\nu_b = e^c_\nu e_a^\mu (\partial_\mu e^\nu_b + e^\lambda_b \Gamma^\nu_{\mu\lambda}),
\end{equation}
and it can also be shown that metricity imposes $\Gamma_{abc} = -\Gamma_{cba}$.

Finally, let us note that it is possible to define a covariant derivative which acts in a covariant way both with respect to curved spacetime (as $\nabla$ does) and to the Lorentz frame (as $\mathfrak{D}$ does), which we denote $\mathscr{D}$. This naturally involves the connections both with curved and flat indices. For instance, it acts on the vierbein as
\begin{equation}
    \mathscr{D}_\nu e^\mu_b = e^a_\nu \partial_a e^\mu_b + \Gamma^\mu_{\lambda\nu} e^\lambda_b - \Gamma^c{}_\nu{}_b e^\mu_c,
\end{equation}
which can be shown to vanish $\mathscr{D}_\nu e^\mu_b = \mathscr{D}_\nu e^b_\mu = 0$ due to metricity of the connections.

With all of this set, we can now tackle the derivation of the TDiff version of the Dirac equation in curved spacetime from the action \eqref{eq:dirac_action}. Let us start by integrating by parts the term with $D_\mu\bar\Psi$, as one usually does in flat spacetime. After expanding it, it has the following form
\begin{equation}
    S_{\bar\Psi} = -\frac{i}{2}\int \dd[4]{x} f(g) (D_\mu \bar \Psi) \gamma^\mu \Psi = -\frac{i}{2} \int \dd[4]{x} f(g) (\partial_\mu\bar \Psi - \bar\Psi \Gamma_\mu -iqA_\mu\bar\Psi) \gamma^\mu \Psi.
\end{equation}

We now integrate the first term by parts, expand derivatives and collect the terms in a convenient way to write the following
\begin{equation}
    S_{\bar\Psi} = \frac{i}{2} \int \dd[4]{x} \left( f \bar\Psi \gamma^\mu D_\mu \Psi + \bar\Psi \left( \partial_\mu (f \gamma^\mu) - f [\gamma^\mu, \Gamma_\mu] \right) \Psi \right).
\end{equation}

In the Diff case with $f=\sqrt{g}$ the second term vanishes. However, we show in what follows that this is no longer the case in the TDiff case. Let us compute the commutator of the connection with the curved gamma
\begin{equation}
    [\gamma^\mu, \Gamma_\mu] = \frac{1}{8}e^\mu_c \Gamma^a{}_\mu{}^b \left[ \gamma^c, [\gamma_a, \gamma_b]\right].
\end{equation}

By expanding the commutator, applying the Chisholm identity to the three-gamma products
\begin{equation}
    \gamma^a \gamma^b \gamma^c =  \eta^{ab} \gamma^c + \eta^{bc} \gamma^a - \eta^{ac} \gamma^b + i \epsilon^{abcd} \gamma^d \gamma^5,
\end{equation}
where $\gamma^5 = i \gamma^0 \gamma^1 \gamma^2 \gamma^3$, and performing some Dirac algebra, one can get the following compact expression
\begin{equation}
    [\gamma^\mu, \Gamma_\mu] = - \Gamma^a{}_{cb} \eta^{cb} \gamma_a = \nabla_\mu \gamma^\mu,
\end{equation}
and with this result
\begin{equation}
    \partial_\mu (f \gamma^\mu) - f [\gamma^\mu, \Gamma_\mu] = \gamma^\mu \partial_\mu f - f \Gamma^\mu_{\mu\lambda} \gamma^\lambda = \sqrt{g} \gamma^\mu \partial_\mu\left(\frac{f}{\sqrt{g}}\right),
\end{equation}
where we have used the identity $\Gamma^\mu_{\mu\lambda} = \frac{1}{\sqrt{g}} \partial_\lambda \sqrt{g}$. This term obviously vanishes in the Diff-invariant case $f(g)=\sqrt{g}$ but does not otherwise, so the action, written in its most common form, acquires an additional term
\begin{equation}
    S_D = \int \dd[4]{x} f(g) \bar\Psi \left[i\gamma^\mu D_\mu + \frac{i}{2}\frac{\sqrt{g}}{f(g)} \gamma^\mu \partial_\mu\left(\frac{f}{\sqrt{g}}\right) - m \right] \Psi.
\end{equation}

The Dirac equation is easily obtained by performing variations of this action with respect to $\bar\Psi$
\begin{equation}\label{eq:dirac_equation}
    \left[i\gamma^\mu D_\mu + \frac{i}{2}\frac{\sqrt{g}}{f(g)} \gamma^\mu \partial_\mu\left(\frac{f}{\sqrt{g}}\right) - m \right] \Psi = 0.
\end{equation}

The equation of motion for the Dirac conjugate can be obtained by integrating by parts the other term in the action, or equivalently, by taking the adjoint of this equation of motion. Taking into account that $\gamma^\mu{}^\dagger = \gamma^0 \gamma^\mu \gamma^0$ and $\gamma^0\gamma^0=1$, where these $\gamma^0$ feature a Lorentz index, one can do
\begin{equation}
    \Gamma_\mu^\dagger = \frac{1}{8} \Gamma^a{}_\mu{}^b [\gamma_a, \gamma_b]^\dagger = -\gamma^0 \Gamma_\mu \gamma^0.
\end{equation}

With this, it is immediate to take the conjugate and write
\begin{equation}
    i D_\mu \bar\Psi \gamma^\mu + \frac{i}{2}\frac{\sqrt{g}}{f(g)} \partial_\mu\left(\frac{f}{\sqrt{g}}\right) \bar\Psi \gamma^\mu + m \bar\Psi = 0.
\end{equation}

The new term contains derivatives of the metric, which do not affect the dynamics at leading order of the adiabatic expansion. In other words, our Dirac equation is equivalent to the usual one in the geometric optics approximation, which we will see in what follows. 

Before doing so, one can easily check that the Dirac action is invariant under global U(1) transformations of the spinor field $\Psi \to e^{i\alpha} \Psi$, which as expected implies that the Noether current
\begin{equation}
    j^\mu_N = \frac{f(g)}{\sqrt{g}}\bar \Psi \gamma^\mu \Psi,
\end{equation}
is a vector under TDiff transformations that satisfies
\begin{equation}
     \nabla_\mu \left[ \frac{f(g)}{\sqrt{g}} \bar \Psi \gamma^\mu \Psi\right] = \mathscr{D}_\mu \left[ \frac{f(g)}{\sqrt{g}} \bar \Psi \gamma^\mu \Psi\right] = 0,
\end{equation}
where in the first step we have used that the term in square brackets is a scalar with respect to the covariant derivative $\mathfrak{D}$. Using Leibniz's rule, the equations of motion and changing $\mathscr{D}_\mu \Psi = \mathfrak{D}_\mu \Psi$ when necessary, it is immediate to check the conservation of this current.

Let us examine the Dirac equation in the geometric optics approximation. We write the Dirac field as a rapidly oscillating exponential times a slowly varying amplitude
\begin{equation}
    \Psi (x) = \left(\psi_0 + \psi_1\epsilon + \dots \right) e^{i\theta(x)/ \epsilon},
\end{equation}
where $\epsilon$ is a dummy power-counting parameter. We also define $k_\mu = \partial_\mu \theta(x)$, so that the amplitudes depend both on this momentum and spatial position $\psi_n = \psi_n(x,\partial\theta)$. 

Following an analogy with a semi-classical analysis \cite{Oancea:2022utx}, where the parameter $\epsilon$ would play the role of $\hbar$, every term originating from the purely kinetic part of the action must carry an additional $\epsilon$. Thus, we write
\begin{equation}
    \left[i \gamma^\mu (\epsilon \mathfrak{D}_\mu + iqA_\mu) + \frac{i\epsilon}{2}\frac{\sqrt{g}}{f(g)} \gamma^\mu \partial_\mu\left(\frac{f}{\sqrt{g}}\right) - m \right] \Psi = 0.
\end{equation}
and similarly for its Dirac conjugate. Now it is immediate to realise that the new term, coming from breaking down to TDiff invariance, does not contribute at leading order, as one would expect. The equations of motion and conservation equation at leading order $\mathcal{O}(\epsilon^0)$ are given by
\begin{align}
    (\gamma^\mu p_\mu + m) \psi_0 &= 0,\\
    \bar\psi_0 (\gamma^\mu p_\mu + m) &= 0,\\
    \partial_\mu (f \bar\psi_0 \gamma^\mu \psi_0) &= 0,
\end{align}
with $p_\mu = k_\mu + q A_\mu$. With the aid of the equations of motion, we can write the current as
\begin{equation}
    \bar\psi_0 \gamma^\mu \psi_0 = -\frac{1}{m} p^\mu \bar\psi_0 \psi_0,
\end{equation}
where we have used the anticommutation relations of the gamma matrices as well. With this, the conservation equation can be written 
\begin{equation}
    \nabla_\mu\left( \frac{f}{\sqrt{g}} \bar\psi_0 \psi_0 p^\mu \right) = 0,
\end{equation}
which features the usual $f/\sqrt{g}$ factor that goes along currents. On the other hand, by multiplying both equations of motion together, we can write
\begin{equation}
    \bar\psi_0 \left[ (p_\mu \gamma^\mu)^2 + 2m p_\mu \gamma^\mu + m^2 \right] \psi_0 = 0.
\end{equation}
Using the equations of motion and the fact that $(p_\mu \gamma^\mu)^2 = p_\mu p^\mu$, we obtain
\begin{equation}
    p_\mu p^\mu = m^2,
\end{equation}
which is the usual dispersion relation for a particle under the effect of an electromagnetic field. Let us note that, since $k_\mu$ is the gradient of a scalar, then $\nabla_\mu k_\nu = \nabla_\nu k_\mu$, so
\begin{equation}
    \nabla_\mu p_\nu - \nabla_\nu p_\mu = q (\nabla_\mu A_\nu - \nabla_\nu A_\mu) = q F_{\mu\nu}.
\end{equation}

This allows us to differentiate the dispersion relation and obtain
\begin{equation}
    p^\mu \nabla_\mu p_\nu = q\, p^\mu F_{\mu\nu},
\end{equation}
which is the standard Lorentz-Dirac equation. In this equation, we observe that the Lorentz force in the TDiff case is unchanged with respect to the Diff expression provided $f_D(g)=f(g)$. In Appendix \ref{sec:lorentz}, we present an alternative derivation of the TDiff-invariant Lorentz force law, in which the starting point is the action of a massive point-particle coupled to the electromagnetic field.

\section{Covariant quantization}\label{sec:quantization}

We consider the quantization of the gauge field by extending the usual covariant quantization approach \cite{Itzykson:1980rh, Pfenning:2001wx, BeltranJimenez:2009oao} to the TDiff case. It is well known that the Lorentz gauge condition $\nabla_\mu A^\mu=0 $ cannot be imposed consistently at the operator level but only by restricting the physical Hilbert space of
the theory by means of the so-called Gupta-Bleuler condition. This formalism  requires to modify the action for the vector field as
\begin{equation}\label{eq:gupta_bleuler_action}
    S_A =  \int \dd[4]{x} f(g) \left(-\frac{1}{4}F_{\mu\nu} F^{\mu\nu}- \frac{\xi}{2}(\nabla_\mu A^\mu)^2\right).
\end{equation}
The corresponding equations of motion in vacuum now read
\begin{align}
    \nabla_\mu\left[ \frac{f(g)}{\sqrt{g}} F^{\mu\nu} \right]+\xi \nabla^\nu\left[\frac{f(g)}{\sqrt{g}}\nabla_\mu A^\mu\right]=0 \label{Afree}
\end{align}

We need to find a complete set of modes that solve these equations of motion, so we proceed as usual, performing an expansion of the vector field in modes, in terms of creation and annihilation operators
\begin{equation}\label{eq:vf_modes}
    A_\mu(x) = \int \frac{\dd[3]{\vb{k}}}{(2\pi)^{3/2}} \sum_\lambda \left[ a_{k\lambda} A_{\mu,k\lambda}(x) + a_{k\lambda}^\dagger A_{\mu,k\lambda}^*(x) \right],
\end{equation}
where the sum in polarizations spans four polarizations $\lambda = 0,1,2,3$, only two of which are physical and the field satisfies the Gupta-Bleuler condition
\begin{align}
    \nabla^\mu A_\mu^{(+)}\vert \psi\rangle=0,
\end{align}
where $\vert \psi\rangle$ is a physical state and
$A_\mu^{(+)}$ is the positive frequency part of the field operator. This condition ensures that on physical states $\langle \psi\vert\nabla^\mu A_\mu\vert \psi\rangle=0$. 

The modes of the vector field are chosen to be orthonormal with respect to the inner product, which is defined the usual way \cite{BeltranJimenez:2009oao}:
    \begin{equation}\label{eq:inner_product_covariant}
    (A_{k\lambda}, A_{k'\lambda'}) = i \int_\Sigma \dd{\Sigma_\mu} \left[A_{\nu,k\lambda}^* \Pi^{\mu\nu}_{k'\lambda'} - \Pi_{k\lambda}^{*\mu\nu} A_{\nu,k'\lambda'} \right] = -(2\pi)^3 \eta_{\lambda\lambda'} \delta^{(3)} (\vb{k}-\vb{k'}),
\end{equation}
    where $\dd{\Sigma_\mu} = n_\mu \dd{\Sigma} = n_\mu \sqrt{g_\Sigma} \dd[3]{x}$ is the volume element on the spatial Cauchy hypersurface $\Sigma$, with $n_\mu$ a unit vector normal to the hypersurface $\Sigma$ and $g_\Sigma$ the (absolute value of the) determinant of the metric induced on $\Sigma$ and $\eta_{\lambda\lambda'}=\diag(1,-1,-1,-1)$. In particular, if $n_\mu$ is purely timelike, then $n_\mu = ((g^{00})^{-1/2}, \vb{0})$, and if the metric tensor is diagonal, then $n_\mu = (\sqrt{g_{00}}, \vb{0})$ and $\dd{\Sigma_\mu} = \sqrt{g}(\dd[3]{\vb{x}}, \vb{0})$. 

The generalised conjugate momentum $\Pi^{\mu\nu}$ is defined as
\begin{equation}
    \delta S_A = \int \dd[4]{x} \sqrt{g}\, \Pi^{\mu\nu} \delta(\partial_\mu A_\nu),
\end{equation}
which for our action \eqref{eq:gupta_bleuler_action}, takes the following form
\begin{equation}
    \Pi^{\mu\nu} = - \frac{f(g)}{\sqrt{g}} \left( F^{\mu\nu} + \xi g^{\mu\nu} \nabla_\alpha A^\alpha \right).
\end{equation}

The inner product \eqref{eq:inner_product_covariant} does not depend on the particular choice of hypersurface $\Sigma$, as the current
\begin{equation}
    J^\mu = A_{1\nu}^* \Pi^{\mu\nu}_2 - \Pi_1^{*\mu\nu} A_{2\nu}
\end{equation}
is conserved $\nabla_\mu J^\mu=0$ under the equations of motion, with $A_{1\nu}$ and $A_{2\nu}$ solutions of the free equations of motion \eqref{Afree} and $\Pi_1^{\mu\nu}$ and $\Pi_2^{\mu\nu}$ the corresponding momenta. By defining the momentum projected onto the normal direction $n_\mu$ to the spatial hypersurface $\Sigma$,
\begin{equation}
    \pi^\mu = n_\lambda \Pi^{\lambda \mu},
\end{equation}
we can now use the equal-time commutation relations
\begin{subequations}
\begin{align}
    \left[A_\mu(\tau,\vb{x}), A_\nu(\tau, \vb{x'}) \right] &= 0,\label{eq:commutation_AA}\\
    \left[\pi^{\mu}(\tau,\vb{x}), \pi^{\nu}(\tau, \vb{x'}) \right] &= 0, \label{eq:commutation_PiPi}\\
    \left[A_\mu(\tau,\vb{x}), \pi^{\nu}(\tau, \vb{x'}) \right] &= i \frac{\delta_\mu{}^\nu}{\sqrt{g_\Sigma}} \delta^{(3)}(\vb{x}-\vb{x'})\label{eq:commutation_APi}
\end{align}
\end{subequations}
to obtain the commutation relations between creation and annihilation operators. 

Inserting the expansion of the vector field \eqref{eq:vf_modes} into the first commutation relation \eqref{eq:commutation_AA}, we have
\begin{multline}
    \left[A_\mu(\tau,\vb{x}), A_\nu(\tau, \vb{x'}) \right] = \int\frac{\dd[3]{\vb{k}} \dd[3]\vb{k'}}{(2\pi)^3}\sum_{\lambda\lambda'} \left( [a_{k\lambda},a_{k'\lambda'}] A_{\mu,k\lambda} A_{\nu,k'\lambda'} + [a^\dagger_{k\lambda},a^\dagger_{k'\lambda'}] A^*_{\mu,k\lambda} A^*_{\nu,k'\lambda'} \right.\\
    \left.+ [a_{k\lambda},a^\dagger_{k'\lambda'}] 
     \left(A_{\mu,k\lambda} A_{\nu,k'\lambda'}^* - A^*_{\mu,k'\lambda'} A_{\nu,k\lambda}\right)\right) = 0.
\end{multline}
This, and the same can be obtained through \eqref{eq:commutation_PiPi}, necessarily implies 
\begin{equation}
    [a_{k\lambda},a_{k'\lambda'}] = [a^\dagger_{k\lambda},a^\dagger_{k'\lambda'}] = 0,
\end{equation}
as well as $[a_{k\lambda},a^\dagger_{k'\lambda'}] = f(\vb{k},\lambda) \delta^{(3)}(\vb{k}-\vb{k'}) \delta_{\lambda \lambda'}$, so that the term multiplying this commutator can vanish after using both deltas to eliminate a sum and an integral. The expression for $f(\vb{k},\lambda)$ can be obtained from \eqref{eq:commutation_APi}, which reads
\begin{equation}
     \int \frac{\dd[3]{\vb{k}} \dd[3]\vb{k'}}{(2\pi)^3} n_\lambda \sum_{\lambda\lambda'} [a_{k\lambda}, a^\dagger_{k'\lambda'}] \left( A_{\mu, k\lambda} \Pi_{k'\lambda'}^{*\lambda \nu} - \Pi_{k\lambda}^{\lambda \nu} A^*_{\mu, k'\lambda'} \right) = i \frac{\delta_\mu{}^\nu}{\sqrt{g_\Sigma}} \delta^{(3)}(\vb{x}-\vb{x'}).
\end{equation}
We now multiply by $\sqrt{g_\Sigma}$, integrate over space and contract the free indices to find
\begin{equation}
     \int \frac{\dd[3]{\vb{k}} \dd[3]\vb{k'}}{(2\pi)^3} \sum_{\lambda\lambda'} [a_{k\lambda}, a^\dagger_{k'\lambda'}] \int \dd[3]{x} \sqrt{g_\Sigma} n_\lambda \left( A_{\mu, k\lambda} \Pi_{k'\lambda'}^{*\lambda \mu} - \Pi_{k\lambda}^{\lambda \mu} A^*_{\mu, k'\lambda'} \right) = 4i.
\end{equation}

Finally, by using the normalization condition \eqref{eq:inner_product_covariant} to replace the spatial integral, the commutator needs to be
\begin{equation}
    [a_{k\lambda}, a^\dagger_{k'\lambda'}] = -\eta_{\lambda \lambda'} \delta^{(3)}(\vb{k}-\vb{k'}), \; \text{with} \; \lambda,\lambda'=0,1,2,3
\end{equation}

\subsection*{Geometric optics limit}

Let us now use the inner product we have just defined to find the normalization of the vector field modes in the geometric optics approximation. In order not to overcrowd the equations, allow us to write the modes of the field now as
\begin{equation}
    A_{\mu,k\lambda}(x) = U_{\mu,k\lambda} (x) e^{i\theta_{k\lambda}(x)},
\end{equation}
where we have dropped the dummy expansion parameter and gathered the whole amplitude into a single object. In terms of the field strength, the scalar product can be written as
\begin{align}
    (A_{k\lambda}, A_{k'\lambda'}) = -i \int \dd[3]{\vb{x}} \sqrt{g_\Sigma} n_\mu \frac{f}{\sqrt{g}} \left[A_{\nu,k\lambda}^* F^{\mu\nu}_{k'\lambda'} - F_{k\lambda}^{*\mu\nu} A_{\nu,k'\lambda'} + \xi g^{\mu\nu} (A^*_{\nu,k\lambda} \nabla_\alpha A^\alpha_{k'\lambda'} - \nabla_\alpha A^{*\alpha}_{k\lambda} A_{\nu,k'\lambda'} ) \right].
\end{align}

The field strength in terms of the modes is
\begin{equation}
    F_{k\lambda}^{\mu\nu} = g^{\mu\rho} g^{\nu\sigma} (ik_\rho U_{\sigma,k\lambda} - ik_\sigma U_{\rho,k\lambda} + \partial_\rho U_{\sigma,k\lambda} + \partial_\sigma U_{\rho,k\lambda}) e^{i\theta_{k\lambda}}.
\end{equation}

At leading order, terms involving derivatives of the amplitude $\partial_\mu U_\nu$ do not contribute Introducing these two expressions into the inner product \eqref{eq:inner_product_covariant}, with $k'\lambda' = k\lambda$, we get the following expression at leading order
\begin{equation}
    (A_{k\lambda}, A_{k\lambda}) = 2 \sqrt{g_\Sigma} \frac{f}{\sqrt{g}}(2\pi)^3 \delta^{(3)}(\vb{0}) n_\mu  \left[ k^\mu U^*_{\sigma,k\lambda} U^{\sigma}_{k\lambda} + (\xi - 1) \Re(k^\sigma U_{\mu,k\lambda} U^{*}_{\sigma,k\lambda}) \right].
\end{equation}

In the Feynman gauge $\xi = 1$, the last term vanishes, and we can impose \eqref{eq:inner_product_covariant} to obtain the normalization of the vector field
\begin{equation}
    U^*_{\rho,k\lambda} U^\rho_{k\lambda} = \frac{-\eta_{\lambda\lambda}}{2f n_\mu k^\mu} \sqrt{\frac{g}{g_\Sigma}}.
\end{equation}

As a result, the vector field mode at leading order in geometric optics can be finally written as
\begin{equation}\label{eq:A_geometric_normalised}
    A_{\mu,k\lambda}(x) = \left(\frac{g}{g_\Sigma}\right)^{1/4} \frac{1}{\sqrt{2f |n_\mu k^\mu|}} u_{\mu,k\lambda} e^{i\theta(x)},
\end{equation}
with $u_{\mu,k\lambda}$ a polarization vector that satisfies $k_\mu u^\mu_{k\lambda} = 0$ and $u^*_{\mu,k\lambda} u^\mu_{k\lambda} = \eta_{\lambda\lambda}$.

\section{TDiff vector fields in cosmological backgrounds}\label{sec:homogeneous}

Let us turn our attention to a cosmological scenario by examining the dynamics of vector fields in a homogeneous and isotropic expanding universe. The line element is given by the flat Robertson-Walker metric
\begin{equation}\label{eq:RWab}
    \dd{s}^2 = b^2(\tau) \dd{\tau}^2 - a^2(\tau) \dd{\vb{x}}^2,
\end{equation}
where $b(\tau)$, $a(\tau)$ are the lapse function and scale factor, respectively. Note that due to not having Diff invariance, we cannot start by changing the time coordinate to the usual cosmological time $\dd{t} = b(\tau) \dd{\tau}$ \cite{Maroto:2023toq}. 

\subsection{Homogeneous fields}
Let us start by considering the simple case 
of homogeneous vector fields $A_\mu(\tau) = (A_0(\tau), \vb{A}(\tau))$. The equations of motion \eqref{eq:general_eom}   are given in this case by
\begin{equation}\label{eq:eom_A_hom}
    \vb{A}'' + \left[ (6f_1 - 2)\frac{a'}{a}+(2f_1-2)\frac{b'}{b}\right]\vb{A}' = 0,
\end{equation}
with $A_0(\tau)$ unconstrained, as it does not appear in the field strength tensor, so it can be determined via gauge fixing. Here and everywhere except on the function $f(g)$, the primes represent derivatives with respect to the time coordinate $'=\dd/\dd{\tau}$. In principle, this equation cannot be integrated unless we can solve Einstein's equations for $a(\tau)$ and $b(\tau)$, even if we assume a specific $f(g)$.

The stress-energy tensor components for this configuration (the energy density $\rho$ and pressure in each direction $p_i$) are given by
\begin{align}
    \rho &= T^0{}_0 = \frac{f(g)}{\sqrt{g}}(1-f_1)\frac{|\vb{A'}|^2}{a^2 b^2},\\
    -p_i &= T^i{}_i = \frac{f(g)}{\sqrt{g}}\frac{1}{a^2b^2}\left( A_i'^2 - f_1 |\vb{A'}|^2\right)\quad\text{(no sum over $i$)},
\end{align}
with the off-diagonal elements $T^0{}_{i} = T^i{}_0 = T^i{}_j (i\neq j)=0$. Due to the nature of the vector field, we have a configuration with anisotropic pressures, which in general do not need to be equal. However, we can compute a mean equation of state as
\begin{equation}
    \bar w = \frac{\bar p}{\rho} = \frac{\frac{1}{3}\sum_i p_i}{\rho} = \frac{f_1 - \frac{1}{3}}{1-f_1},
\end{equation}
which in the Diff case $f_1 = 1/2$ equals $\bar w = 1/3$, as expected for a free vector field, which behaves as radiation.

One of the main differences of TDiff-invariant field theories is that the stress-energy conservation equations are not automatically fulfilled under solutions to the field's equations of motion. Although the stress-energy tensor is conserved on solutions of the Einstein equations, the non-linearity and complexity of Einstein's equations make it more practical to use conservation equations in order to obtain constraints involving the metric components and their derivatives. This allows us to determine some functions in the metric that we have not been able to fix previously, due to having one less gauge degree of freedom as a result of having a reduced symmetry group. In particular, this is the case of the lapse function $b(\tau)$ in the TDiff-invariant RW metric, as time dilations are not TDiff.

Let us examine now the conservation equations. Explicitly, they are given by
\begin{subequations}
\begin{gather}
    \partial_\mu T^\mu{}_0 + 3\frac{a'}{a}T^0{}_0 - \frac{a'}{a} T^i{}_i = 0,\\
    \partial_\mu T^\mu{}_i +\left(2\frac{a'}{a}+\frac{b'}{b}\right) T^0{}_i - \frac{aa'}{b^2} T^i{}_0 = 0.
\end{gather}    
\end{subequations}

In the case of a homogeneous vector field, the second conservation equation is identically zero, whereas the first one yields the following equation
\begin{equation}\label{eq:pre_cons_hom}
    (2-2f_1) \vb{A''}\cdot\vb{A'} + |\vb{A'}|^2\left[(-3+11f_1-6f_1^2-6f_2)\frac{a'}{a} + (-3+5f_1-2f_1^2 - 2f_2)\frac{b'}{b}\right] = 0.
\end{equation}

We can eliminate the dependence on the vector field by taking the scalar product of the field equation of motion \eqref{eq:eom_A_hom} with $\vb{A'}$ and substituting it into \eqref{eq:pre_cons_hom}, so we have
\begin{equation}\label{eq:electric_condition}
    (1- 5f_1 + 6f_1^2 - 6f_2)\frac{a'}{a} + (1 - 3f_1 + 2f_1^2 -2f_2)\frac{b'}{b}=0.
\end{equation}
We will refer to this constraint as the ``electric condition'' since, as we will show below, it is the constraint that having an electric field imposes onto the spacetime. In particular, it allows to obtain $b(\tau)$ from $a(\tau)$ for a given $f(g)$. Notice that, as expected, in the Diff case with $f_1=1/2$ and 
$f_2=0$ the equation is trivially satisfied for any $b(\tau)$. 
This constraint can be inserted into the equations of motion \eqref{eq:eom_A_hom}, which now become
\begin{equation}
    \vb{A}'' - \frac{8f_2}{1 -3f_1 +2f_1^2 -2f_2} \frac{a'}{a} \vb{A}' = 0.
\end{equation}
Note that, for a $f(g)$ so that $f_2=0$, such as a power law, the equation of motion is simply $\vb{A''}=0$, so the vector potential evolves linearly with time $\vb{A}\propto\tau$ and $F_{0i}=\mathrm{const}$. If we think in terms of electric $\vb{E}$ and magnetic $\vb{B}$ fields, defined as
\begin{equation}\label{eq:def_EB}
    E_i = F_{0i}, \quad B_i = \frac{1}{2}\epsilon_{ijk} F_{jk},
\end{equation}
this particular configuration of a homogeneous vector field corresponds to a constant electric field, as we had anticipated.

In particular, for a power-law $f(g)=g^\alpha$, the explicit dependence between $a$ and $b$ can be extracted by integrating \eqref{eq:electric_condition}, yielding
\begin{equation}\label{eq:electric_relation}
    b(\tau) \propto a^{\frac{1-3\alpha}{\alpha - 1}}(\tau).
\end{equation}
which implies 
\begin{align}
    g\propto a^{4/(1-\alpha)}
\end{align}

If we insert \eqref{eq:electric_relation} into the equation of motion \eqref{eq:eom_A_hom}, the term in square brackets vanishes, as expected, since as mentioned before in this case the equation reduces to $\vb{A''}=0$. Taking into account this time dependence, both the energy density and pressure evolve with the scale factor as
\begin{equation}
    \rho, p_i \propto a^{-2/(1-\alpha)},
\end{equation}
which in the Diff invariant case ($\alpha=1/2$) reads $\rho\propto a^{-4}$, as expected for a radiation component.

\subsubsection{Homogeneous magnetic field}
We work on a configuration with a constant magnetic field now. Taking into account the definition of the magnetic field \eqref{eq:def_EB}, we need a vector potential that takes the following form in order to obtain a homogeneous configuration:
\begin{equation}
    A_i = \frac{1}{2}\epsilon_{ijk} x^j B_k,
\end{equation}
with a constant $\vb{B}$, because in any other case, a non-homogeneous electric field would arise as well. As such, in order to have only a magnetic field, it is forced to be constant by construction. Since the field strength $F_{ij}=\mathrm{const.}$, with all other components zero, it is easy to see that it satisfies the equations of motion \eqref{eq:general_eom}.

The non-zero components of the stress-energy tensor now are
\begin{align}
    \rho &= T^0{}_0 = \frac{f(g)}{\sqrt{g}}\frac{f_1}{a^4}\vb{B}^2,\\
    -p_i &= T^i{}_i = \frac{f(g)}{\sqrt{g}}\frac{1}{a^4}[(f_1 - 1) \vb{B}^2 + B_i^2]\quad\text{(no sum over $i$)},\\
    T^i{}_j &= \frac{f(g)}{\sqrt{g}} \frac{1}{a^4} \left((f_1-1) \vb{B}^2 \delta^i{}_j + B_i B_j\right), 
\end{align}
where we have used that $F_{ij} F_{ij} = 2\vb{B}^2$. The average equation of state is
\begin{equation}
    \bar w = \frac{\bar p}{\rho} = \frac{f_1}{3 (2 - 3f_1)},
\end{equation}
which again yields $\bar w = 1/3$ for the Diff-invariant case.

The conservation of the stress-energy tensor together with the field equations of motion impose now the ``magnetic condition'', namely
\begin{equation}\label{eq:magnetic_condition}
    \frac{a'}{a}\left(2-7f_1+6f_1^2 +6f_2\right) + \frac{b'}{b}(-f_1 +2f_1^2 + 2f_2)=0.
\end{equation}
Notice once more that in the Diff case $f_1=1/2$ and $f_2=0$ the condition is trivially satisfied. 
 In the particular case of a power-law $f(g)=g^\alpha$, this condition is solved by the following relation between $b$ and $a$
\begin{equation}\label{eq:magnetic_relation}
    b(\tau) \propto a^\frac{2-3\alpha}{\alpha}(\tau).
\end{equation}
which implies 
\begin{align}
    g\propto a^{4/\alpha}
\end{align}

Interestingly, both the condition and the relationship between $b$ and $a$ are different from those in the 
purely electric case (\ref{eq:electric_condition},~\ref{eq:electric_relation}), which translates into a different spacetime depending on the vector field configuration that it hosts. Notice however that these conditions approach the same values as $\alpha\to 1/2$.

The energy density and pressures now scale as
\begin{equation}
    \rho, p_i \propto a^{-2/\alpha}.
\end{equation}
with the same radiation-like behaviour $\rho\propto a^{-4}$ in the Diff-invariant case. Notice however that for $\alpha>1/2$ the magnetic energy density dilutes more slowly than standard radiation, which could provide a mechanism for amplification of primordial magnetic fields \cite{Turner:1987bw, Durrer:2013pga}. As a matter of fact, unlike other amplification mechanisms, it could operate on intergalactic magnetic fields as well, which would allow to have sizeable magnetic fields today without requiring a large primordial seed.

Following these different results, we wonder whether it is possible to have both TDiff-invariant electric and magnetic fields in a Robertson-Walker background while satisfying the corresponding conservation equations. In that case, we should be able to find a more general solution to the conservation equation $b(a)$ that interpolates between the two we have already obtained. 

\subsubsection{General homogenous field}\label{sec:conservation}
Let us look back at the individual electric and magnetic conditions separately for a general $f(g)$.  Since the various $f_n$ that appear in these equations are in general a function of the metric determinant $g$, it is useful to substitute $b=b(g,a)$, which for our matter the expression
\begin{equation}
    \frac{b'}{b} = \frac{1}{2}\frac{g'}{g}-3\frac{a'}{a}
\end{equation}
will suffice. Inserting this into the electric condition \eqref{eq:electric_condition}, we obtain
\begin{equation}
    \frac{a'}{a}\left(-2+4f_1\right) + \frac{1}{2}\frac{g'}{g}(1-3f_1+2f_1^2-2f_2)=0.
\end{equation}

By making $t=\log g$, $F=\log f$ and integrating we get to
\begin{equation}
   \log a = \frac{1}{4}\int \dd{t}\frac{1 - 2\dot{F} +2\dot{F}^2 -2\ddot{F}}{1-2\dot{F}}, 
\end{equation}
where the overdots denote derivatives with respect to $t$. After performing the integration it becomes the following condition
\begin{equation}
    C_E a^4 = \frac{g}{f}(1-2f_1),
\end{equation}
with $C_E$ a constant.

Working in a similar manner, the magnetic condition \eqref{eq:magnetic_condition} can be written as 
\begin{equation}
    \frac{a'}{a}(1-2f_1) + \frac{1}{4}\frac{g'}{g}(-f_1+2f_1^2 +2f_2) = 0,
\end{equation}
which can be solved to
\begin{equation}
    C_Ba^4 = f (1-2f_1),
\end{equation}
with $C_B$ a constant.

Again, note how the electric and magnetic conditions are different. It is also easy to realise that these constraints are automatically fulfilled when $f=\sqrt{g}$, as well as equivalent to (\ref{eq:electric_relation}, \ref{eq:magnetic_relation}) when introducing $f(g)=g^\alpha$.

In order to examine whether a consistent conservation equation with an electromagnetic configuration can be obtained, we can write, without loss of generality, the following expression for the vector potential
\begin{equation}
    A_i(\tau) = \phi_i(\tau) + \frac{1}{2}\epsilon_{ijk} x^j B_k,
\end{equation}
which gives the most general homogeneous configuration, with electric field $E_i(\tau) = \phi_i'(\tau)$ and magnetic field $B_i$. The magnetic field needs to be constant because otherwise a non-homogeneous electric field is invoked, as discussed in the previous section.

Unfortunately, unlike in the purely electric or magnetic case, the conservation equation cannot be obtained for a general $f(g)$, so throughout the rest of this section, we assume a power-law $f(g)=g^\alpha$. The time evolution of the electric field can easily be obtained thanks to the equation of motion \eqref{eq:eom_A_hom}, which solves to
\begin{equation}
    \vb{E}(\tau) = \vb{E_0} a^{2-6\alpha}(\tau) b^{2-2\alpha}(\tau),
\end{equation}
where $\vb{E_0}$ is the value of the electric field when $a=b=1$. In this case, the conservation of the stress-energy tensor imposes the following constraint
\begin{align}
    \frac{E_0^2}{B^2}(a^6 b^2)^{1-2\alpha} \left[\frac{a'}{a}(1-5\alpha+6\alpha^2) + \frac{b'}{b}(1-3\alpha+2\alpha^2)\right] + \left[\frac{a'}{a}\left(1-\frac{7}{2}\alpha +3\alpha^2\right) + \frac{b'}{b}\left(-\frac{\alpha}{2}+\alpha^2\right)\right]=0,
\end{align}
where $E_0 = |\vb{E_0}|$ and $B = |\vb{B}|$. Making the substitution $b = b(g,a)$ as in the previous cases, we obtain
\begin{equation}
    (1-2\alpha)\frac{a'}{a}\left(2 \frac{E_0^2}{B^2}g^{1-2\alpha} - 1\right) = (1-2\alpha)\frac{g'}{g} \left(\frac{1-\alpha}{2}\frac{E_0^2}{B^2}g^{1-2\alpha}-\frac{\alpha}{4}\right).
\end{equation}
Notice that the condition is automatically satisfied in the Diff case $\alpha=1/2$ as expected. For $\alpha\neq 1/2$, we can divide this expression by $1-2\alpha$ and it can be easily integrated to obtain the following relation
\begin{equation}\label{eq:em_condition}
    C_{EM}a^4 = g^\alpha-2g^{1-\alpha}\frac{E_0^2}{B^2}.
\end{equation}

This expression generalizes the electric and magnetic conditions for a power-law $f(g)$. As one could have predicted, it depends on the values of the electric and magnetic fields, and it is compatible with the electric \eqref{eq:electric_condition} and magnetic \eqref{eq:magnetic_condition} conditions in the $E_0\gg B$ and $E_0\ll B$ limits, respectively. Note that this relation works only for homogeneous fields which, speaking in practical terms, means that we will be able to use it for modes in the super-Hubble regime. Sub-Hubble modes, as we will explore in future sections, fall into the geometric optics approximation and behave as in a Diff-invariant theory.

\subsection{Inhomogeneous fields}\label{sec:general}
In order to study more general configurations, we need to consider a vector field $A_\mu(\tau,\vb{x})$ that depends both on time coordinate and spatial position. The equations of motion \eqref{eq:general_eom} now read, in terms of the field strength
\begin{subequations}\label{eq:em_eom}
    \begin{align}
        \partial_i F_{0i} &= 0,\label{eq:em_eom_0}\\
    \partial_0 F_{0i} &+ \left[ (6f_1 - 2)\frac{a'}{a}+(2f_1-2)\frac{b'}{b}\right] F_{0i} = \frac{b^2}{a^2}\partial_j F_{ji}.\label{eq:em_eom_i}
    \end{align}
\end{subequations}

These equations cannot be solved unless we invoke a particular ansatz for either the field strength or the vector field, which we have already done in the previous section.

The stress-energy tensor has the following components:
\begin{subequations}
\begin{align}
    T^0{}_0 &= \rho = \frac{f}{\sqrt{g}}\left[\frac{1-f_1}{a^2b^2}F_{0i} F_{0i} + \frac{f_1}{2a^4} F_{ij} F_{ij}\right] ,\label{eq:T00_general}\\
    T^0{}_i &= \frac{f}{\sqrt{g}}\frac{1}{a^2b^2} F_{0j} F_{ij},\\
    T^i{}_0 &= -\frac{f}{\sqrt{g}} \frac{1}{a^4} F_{0j} F_{ij},\\
    T^i{}_j &= \frac{f}{\sqrt{g}}\left[\frac{1}{2}f_1\left(\frac{1}{a^4}F_{kl} F_{kl}-\frac{2}{a^2b^2} F_{0k} F_{0k}\right) \delta^i{}_j + \frac{1}{a^2b^2} F_{i0} F_{j0} - \frac{1}{a^4} F_{ik} F_{jk}\right],\\
    T^i{}_i &=   \tr( T^i{}_j ) = -3 \bar{p} = \frac{f}{\sqrt{g}} \left[\frac{1-3f_1}{a^2b^2} F_{0i}F_{0i} +\frac{\frac{3}{2}f_1-1}{a^4} F_{ij} F_{ij}\right].
\end{align}
\end{subequations}

With these, we can write the conservation equations
\begin{subequations}
\begin{multline}
    2(1-f_1)F_{0i}\partial_0 F_{0i} + \frac{b^2}{a^2} f_1 F_{ij}\partial_0 F_{ij} - \frac{b^2}{a^2}\partial_i (F_{0j}F_{ij}) \\
    + F_{0i} F_{0i} \left[\frac{a'}{a}\left(-3+11f_1-6f_1^2-6f_2\right) + \frac{b'}{b}(-3+5f_1-2f_1^2-2f_2)\right]\\
    + \frac{b^2}{a^2} F_{ij} F_{ij} \left[\frac{a'}{a}\left(-3+11f_1-6f_1^2-6f_2\right) + \frac{b'}{b}(-3+5f_1-2f_1^2-2f_2)\right]=0,
\end{multline}
\begin{multline}
    \partial_0(F_{0j}F_{ij}) + \frac{b^2}{a^2}f_1 F_{kl}\partial_i F_{kl} - 2f_1 F_{0j}\partial_i F_{0j} + \partial_j(F_{i0}F_{j0}) - \frac{b^2}{a^2} \partial_j(F_{ik} F_{jk}) \\
    + F_{0j} F_{ij} \left[\frac{a'}{a}(-2+6f_1) - \frac{b'}{b}(-2+2f_1)\right] = 0.
\end{multline}    
\end{subequations}

After inserting the equations of motion into the conservation equations, we get
\begin{subequations}\label{eq:cons_complete}
\begin{multline}\label{eq:cons_complete_0}
    (1-2f_1)\frac{b^2}{a^2}F_{0j}\partial_i F_{ij} + \frac{b^2}{a^2} f_1 F_{ij}\partial_0 F_{ij} - \frac{b^2}{a^2}F_{ij}\partial_i F_{0j} \\
    + F_{0i} F_{0i} \left[\frac{a'}{a}\left(1-5f_1+6f_1^2-6f_2\right) + \frac{b'}{b}(1-3f_1+2f_1^2-2f_2)\right]\\
    + \frac{1}{2}\frac{b^2}{a^2} F_{ij} F_{ij} \left[\frac{a'}{a}\left(2-7f_1+6f_1^2 +6f_2\right) + \frac{b'}{b}(-f_1 +2f_1^2 + 2f_2)\right]=0,
\end{multline}
\begin{equation}\label{eq:cons_complete_i}
    F_{0j}\partial_0 F_{ij} -2f_1 F_{0j}\partial_i F_{0j} + F_{0j}\partial_j F_{0i} + \frac{b^2}{a^2}f_1 F_{jk}\partial_i F_{jk} - \frac{b^2}{a^2}F_{jk}\partial_j F_{ik} = 0.
\end{equation}
\end{subequations}

In these equations, the terms in square brackets vanish if we set $f=\sqrt{g}$, whereas the rest of the terms cancel out with the aid of the Bianchi identities $\partial_{(i} F_{jk)} = 0$, as expected since the conservation of the stress-energy tensor is identical in the Diff-invariant case.

Note that the term in square brackets that multiplies $F_{0i}F_{0i}$ in \eqref{eq:cons_complete_0} is what we dubbed the electric condition \eqref{eq:electric_condition}. As $F_{0i}F_{0i}=\vb{E}^2$, this is the only term that survives when we have a homogeneous electric field. Similarly, $F_{ij}F_{ij} = 2\vb{B}^2$, so we easily identify the term in square brackets that multiplies this combination as the magnetic condition \eqref{eq:magnetic_condition}. 

We can also observe in \eqref{eq:T00_general} that the electric and magnetic parts of TDiff fields gravitate differently. Thus, the electric energy density is proportional to $1-f_1$ whereas the magnetic energy density is proportional to $f_1$, so that they only gravitate in the same proportion in the Diff invariant case.

\subsubsection{Maxwell's equations}

In this section, we take a look at the equations of motion of a free electromagnetic field \eqref{eq:em_eom} written in terms of the electric and magnetic fields \eqref{eq:def_EB}. These read as follows
\begin{subequations}
    \begin{align}
    \nabla\cdot \vb{E} &= 0,\label{eq:gauss_electric}\\
    \frac{b^2}{a^2}\nabla \cross \vb{B} &= \vb{E}' + \left[ (6f_1 - 2)\frac{a'}{a}+(2f_1-2)\frac{b'}{b}\right] \vb{E}, \label{eq:ampere}
\end{align}
\end{subequations}
which correspond to the well-known Gauss's law for the electric field and Ampère's law, respectively, with a slight change due to the Diff-invariance breaking. On top of these, the Bianchi identities are also satisfied by definition of the field strength
\begin{equation}\label{eq:bianchi_f}
    \partial_\mu F_{\alpha\beta} + \partial_\alpha F_{\beta\mu} + \partial_\beta F_{\mu\alpha} = 0.
\end{equation}

By substituting $(\mu,\alpha,\beta) = (0, i, j)$ and contracting with $\epsilon_{ijk}$, we obtain Faraday's law
\begin{equation}\label{eq:faraday}
    \vb{B}' + \nabla\cross\vb{E} = 0,
\end{equation}
whereas doing $(\mu,\alpha,\beta) = (i, j,k)$ and contracting with $\epsilon_{ijk}$, we complete our set of Maxwell equations in vacuum with Gauss's law for the magnetic field
\begin{equation}\label{eq:gauss_magnetic}
    \nabla\cdot\vb{B} = 0.
\end{equation}

For completeness' sake, we also have the conservation equations, which after substituting the equations of motion into them (see Eq. \eqref{eq:cons_complete}) acquire the following form:
\begin{subequations}
    \begin{align}
    (1-2f_1) \left[\vb{E}\cdot(\nabla\cross\vb{B}) + \vb{B}\cdot(\nabla\cross\vb{E})\right] + \frac{a^2}{b^2} \vb{E}^2 &\left[\frac{a'}{a}\left(1-5f_1+6f_1^2-6f_2\right) + \frac{b'}{b}(1-3f_1+2f_1^2-2f_2)\right] \nonumber\\
    + \vb{B}^2 &\left[\frac{a'}{a}\left(2-7f_1+6f_1^2 +6f_2\right) + \frac{b'}{b}(-f_1 +2f_1^2 + 2f_2)\right]=0,
\end{align}
\begin{equation}
    (1-2f_1)\nabla\left( \vb{E}^2-\frac{b^2}{a^2}\vb{B}^2\right) = 0.
\end{equation}
\end{subequations}

Working with the equations in this form is rather complicated, so we might seek to eliminate the electric field from the equations of motion. We can easily do so by taking the curl of Ampère's law \eqref{eq:ampere} and using Faraday's law \eqref{eq:faraday} in it, which leads us to an equation of motion for the magnetic field
\begin{equation}\label{eq:eom_B}
    \vb{B}'' - \frac{b^2}{a^2}\nabla^2\vb{B} + \left[ (6f_1 - 2)\frac{a'}{a}+(2f_1-2)\frac{b'}{b}\right] \vb{B}' = 0.
\end{equation}

Working in Fourier modes, we can analyse the super-Hubble regime ($k (b/a)\ll a'/a, b'/b$), with $k=\vert\vb k\vert$ which features a straight-forward general solution for the magnetic field derivative provided a power-law $f(g)=g^\alpha$
\begin{equation}
    \vb{B}_k'(\tau) \propto a^{2-6\alpha}(\tau) b^{2-2\alpha}(\tau).
\end{equation}
Note, however, that if we require the electric field to have a super-Hubble behaviour as well, Faraday's law \eqref{eq:faraday} necessarily implies $\vb{B}_k'=0$, as we already noted in Sec. \ref{sec:conservation}.

\subsubsection{Sub-Hubble regime}

In order to analyse the sub-Hubble limit, we cannot neglect the first derivative term in Eq. \eqref{eq:eom_B} straight away. Working with a power law $f(g)=g^\alpha$, we perform a change of function $\vb{B}_k(\tau) = a^{\frac{3}{2}-3\alpha}(\tau) b^{\frac{1}{2}-\alpha}(\tau) \hat{\vb{B}}_k(\tau)$ that eliminates the first derivative term. We also perform a change of variable to conformal time $\eta$, $\dd\tau = \frac{a}{b} \dd\eta$, which leaves the equation of motion as follows
\begin{multline}
    \hat{\vb{B}}_k^{(2)} + \left( k^2 + \frac{1}{4}\left[ 3(2\alpha-1) (6\alpha-5) \left(\frac{a^{(1)}}{a}\right)^2 + 6(2\alpha-1)\left((2\alpha-1) \frac{a^{(1)}b^{(1)}}{ab} + \frac{a^{(2)}}{a}\right)\right.\right.\\
    \left.\left.+ (4\alpha(2-\alpha)-3) \left(\frac{b^{(1)}}{b}\right)^2 + 2(1-2\alpha) \frac{b^{(2)}}{b} \right] \right) \hat{\vb{B}}_k  = 0,
\end{multline}
where the superscript in parenthesis $a^{(n)}$ denotes the $n$-th derivative with respect to conformal time $\eta$. In the sub-Hubble limit, the term in square brackets can be ignored and this function admits an easy solution in terms of complex exponentials
\begin{equation}
    \hat{\vb{B}}_k(\eta) = \vb{c}_1 e^{-ik\eta} + \vb{c}_1^* e^{ik\eta},
\end{equation}
where we have imposed that the field is real.

Reverting to the original variables, we have got
\begin{equation}
    {\vb{B}}_k(\tau) = g^{(1-2\alpha)/4}\left(\vb{c}_1 e^{-ik\eta} + \vb{c}_1^* e^{ik\eta}\right),\quad \eta = \int^\tau \dd{s} \frac{b(s)}{a(s)},
\end{equation}
and if we introduce this behaviour into the magnetic part of the energy density (the remaining exponentials can be integrated out thanks to rapid oscillation), we obtain
\begin{equation}
    \langle\rho_B \rangle = \frac{f}{\sqrt{g}}\frac{f_1}{2a^4} \langle F_{ij} F_{ij} \rangle = g^{\alpha-1/2} \frac{\alpha}{a^4} \langle \vb{B}^2 \rangle  =\frac{2\alpha\vert \vb{c}_1 \vert^2}{a^{4}},
\end{equation}
which scales with the expansion exactly as in the Diff-invariant case. Thus, sub-Hubble modes behave as if the Diff-invariance had not been broken down to TDiff, a result that we revisit with the geometric optics approximation in what follows.

The same analysis can be performed in terms of the electric field instead of the magnetic field. If we take the curl of Faraday's law \eqref{eq:faraday} and insert it into the time derivative of Ampère's law \eqref{eq:ampere}, we obtain the following equation of motion for the electric field
\begin{multline}
    \vb{E}'' - \frac{b^2}{a^2} \nabla^2 \vb{E} + \left[ 6f_1 \frac{a'}{a} + 2(f_1-2) \frac{b'}{b} \right] \vb{E}' + \left[ 2 (3f_1 + 18f_2 -1) \frac{a'^2}{a^2} + 2(3 - 3f_1 + 2f_2) \frac{b'^2}{b^2} \right.\\
    \left. - 8(f_1 - 3f_2) \frac{a'}{a} \frac{b'}{b} + 2(3f_1 -1) \frac{a''}{a} + 2(f_1 - 1)\frac{b''}{b} \right] \vb{E} = 0.
\end{multline}

By working in Fourier space, performing a change of function $E_k (\tau) = a^{-3\alpha + \frac{1}{2}}(\tau) b^{-\alpha + \frac{3}{2}}(\tau) \hat E_k(\tau)$ and changing variable to conformal time $\eta$, the equation of motion has the following expression
\begin{multline}
    \hat{\vb{E}}_k^{(2)} + \left( k^2 + \frac{1}{4}\left[ (1-4\alpha^2) \left(\frac{b^{(1)}}{b}\right)^2 - 3(1- 8\alpha + 12\alpha^2) \left(\frac{a^{(1)}}{a}\right)^2 - 6(2\alpha-1)^2 \frac{a^{(1)}}{a}\frac{b^{(1)}}{b} \right.\right.\\
    \left.\left. + 6(2\alpha-1) \frac{a^{(2)}}{a} + 2(2\alpha-1) \frac{b^{(2)}}{b} \right] \right) \hat{\vb{E}}_k  = 0.
\end{multline}

Again, we can neglect the whole term in square brackets in the sub-Hubble regime ($k\gg a^{(1)}/a,b^{(1)}/b$), and obtain a solution for the electric field in terms of complex exponentials
\begin{equation}
    \hat{\vb{E}}_k(\eta) = \vb{c}_2 e^{-ik\eta} + \vb{c}_2^* e^{ik\eta}.
\end{equation}

In terms of the original variables, we have got
\begin{equation}
    \vb{E}_k(\tau) = g^{(1-2\alpha)/4} \frac{b}{a} (\vb{c}_2 e^{-ik\eta} + \vb{c}_2^* e^{ik\eta}),\quad\eta = \int^\tau \dd{s} \frac{b(s)}{a(s)},
\end{equation}
which implies a typical radiation behaviour for the average energy density
\begin{equation}
    \langle\rho_E \rangle = \frac{f}{\sqrt{g}} \frac{1-f_1}{a^2b^2}\langle F_{0i} F_{0i} \rangle = g^{\alpha-1/2}\frac{1-\alpha}{a^2 b^2} \langle \vb{E}^2 \rangle= 2(1-\alpha) \frac{\vert \vb{c}_2 \vert^2}{a^{4}},
\end{equation}
which again scales with the expansion as in the Diff case. 
One last thing to note about these expressions is that, while the scaling is exactly as in the Diff-invariant case, the electric and magnetic fields have different shares of the total energy density, as they are weighed by factors $1-\alpha$ and $\alpha$, respectively. However, Faraday's law \eqref{eq:faraday} imposes
\begin{equation}
    \vb{c}_1 = \hat{\vb{k}} \cross \vb{c}_2
\end{equation}
 on the solutions we have just obtained for the electric and magnetic fields, with $\hat{\vb{k}}=\vb{k}/\vert\vb{k}\vert$  a unit vector in the propagation direction. Since the electric field is transverse, this implies $|\vb{c}_1| = |\vb{c}_2|$, so when computing the total electromagnetic energy density 
\begin{equation}
    \langle\rho_\mathrm{EM} \rangle =   \langle\rho_E\rangle + \langle\rho_B \rangle   =
    2(1-\alpha) \frac{\vert \vb{c}_2 \vert^2}{a^{4}} + 2\frac{\alpha\vert \vb{c}_1 \vert^2}{a^{4}}= \frac{2\vert \vb{c}_1 \vert^2}{a^{4}},
\end{equation}
the dependence on $\alpha$ cancels out and it does not depend on the TDiff function at all.

\subsubsection{Geometric optics approximation}

In an expanding universe, the sub-Hubble regime corresponds in practice to the geometric optics approximation. The ansatz for the field is slightly less general, doing a pure plane wave expansion for the spatial part and separating spatial and temporal dependencies in the rapidly oscillating exponential
\begin{equation}
    A_{\mu,k\lambda}(x) = U_{\mu, k\lambda} (x) e^{i\vb{k}\cdot \vb{x} - i\int^\tau \omega_k(\tau')\dd{\tau'} }.
\end{equation}

We can now use every result we obtained in Sec. \ref{sec:geometric_optics} by identifying
\begin{equation}
    \theta_{k\lambda} (x) = \vb{k}\cdot\vb{x} - \int^\tau \omega_k(\tau')\dd{\tau'}
\end{equation}
and
\begin{equation}
    k_\mu = \partial_\mu \theta = (-\omega_k, \vb{k}).
\end{equation}

The Lorentz gauge condition \eqref{eq:gauge_LO} now reads
\begin{equation}
    a^2\omega_k U_\tau + b^2 \vb{k}\cdot\vb{U} = 0,
\end{equation}
and the equation of motion \eqref{eq:eom_LO} gives us the usual dispersion relation of a massless field
\begin{equation}
    \omega_k^2 = \frac{b^2}{a^2} \vb{k}^2.
\end{equation}

The normalization of the amplitude $U_\mu$ can be obtained either from the next-to-leading order equation or the inner product normalization condition. We use the later method, as we already have a final expression in Eq. \eqref{eq:A_geometric_normalised}. With a RW background, the normal vector has to be $n_\mu = (b, \vb{0})$, so we easily obtain
\begin{equation}\label{eq:A_wkb_normalised}
    A_{\mu, k\lambda}(x) = \sqrt{\frac{b^2}{2f\omega_k}} u_\mu e^{i\vb{k}\cdot \vb{x} - i\int^\tau \omega_k(\tau')\dd{\tau'} },
\end{equation}
with $u_\mu k^\mu = 0$ and $u^*_\mu u^\mu = -1$. Using these expressions for the modes of the vector field in computing the expectation value of the energy density \eqref{eq:T00_general}, one can find $\langle \rho \rangle \propto a^{-4}$. Again, the dynamics in the geometric optics approximation are equivalent to those in a Diff-invariant scenario, so breaking down to TDiff invariance does not disturb the well-settled dynamics on short scales.

\section{Discussion and Conclusions}\label{sec:discussion}

In this work, we have studied the dynamics of Abelian gauge fields which break diffeomorphism invariance down to transverse diffeomorphisms.  We have shown that in the geometric optics approximation, very much as for Diff invariant theories,  the corresponding massless gauge bosons propagate along null geodesics and particle number is conserved. In addition, 
the polarization vectors are orthogonal to the propagation direction and the physical (transverse projection) polarization is parallel transported along the geodesics. We have also studied the coupling to TDiff invariant Dirac spinors.
We conclude that in order to avoid violations of Einstein's Equivalence Principle i.e. either Weak Equivalence Principle violations or violations of Local Position Invariance, the breaking of Diff invariance should be introduced by the same global 
$f(g)$ function for all the different fields. In this case, the standard expressions for the Maxwell and Lorentz-Dirac equations are recovered at leading order in the geometric optics approximation. 

We have also analyzed the contributions to the energy-momentum tensor of the gauge fields. We find that, in general, the breaking of Diff invariance makes the electric and magnetic parts of the vector field gravitate in a different way. In the sub-Hubble regime, we recover the standard radiation-like behaviour of the energy density. However, in the super-Hubble regime, the behaviour is totally different to the Diff case, thus opening up a wide range of possibilities for 
cosmological model building. In particular, for certain $f(g)$, the magnetic energy density could scale more slowly with the expansion, thus effectively amplifying the magnetic fields compared to the standard Diff evolution. This can be understood because breaking down to TDiff invariance also breaks conformal invariance for the gauge field. In fact, breaking the conformal triviality of Maxwell's equations in a RW background is necessary for the production and amplification of cosmological magnetic fields. As a result, the amplification mechanism arises naturally in a TDiff-invariant gauge model. This will be explored in detail in a future work. On the other hand, very much as in the scalar case, the wide range of possibilities for the evolution of the homogeneous vector fields makes these models a useful tool for the construction of models of the dark or the inflationary sectors.

\acknowledgements{We thank Prado Martín-Moruno for useful comments. This work has been supported by the  MICIN (Spain) project PID2022-138263NB-I00 (AEI/FEDER, UE). A.D.M. acknowledges financial support by the MICIU (Spain) through a Formación de Profesorado Universitario (FPU) fellowship FPU18/04599.}

\appendix

\section{Lorentz force}\label{sec:lorentz}
In this Appendix we examine a different derivation of the Lorentz force law for TDiff theories to that presented in Sec. \ref{subsec:dirac}. In the following derivation, we mostly follow that on \cite{Itzykson:1980rh}. 

In the Diff case, the current of a charged point-like particle with charge $q$ following a trajectory $z^\mu(s)$, where $s$ is the length of arc parameter, can be written as
\begin{equation}
    j^\mu (\vb{x},\tau) =  q \int \dd{s} u^\mu \frac{\delta^{(4)}[x-z(s)]}{\sqrt{g}},
\end{equation}
where $u^\mu = \dd{z^\mu}/\dd{s}$ is the four-velocity and we have weighed the Dirac delta to make it a scalar.  The total action in a Diff-invariant theory, including the mass action, would read
\begin{equation}\label{eq:pointaction_currents}
    S = -\int \dd[4]{x} \sqrt{g} \left( \frac{1}{4} F_{\mu\nu} F^{\mu\nu} + j^\mu A_\mu\right) - m \int \dd{s}.
\end{equation}

In turn, the action for the point-like particle of mass $m$ can be written as the four-velocity coupled to a mass current. Since the four-velocity is a unit vector $u_\mu u^\mu = 1$, one can write
\begin{equation}
    \int \dd{s} = \int \dd{s} u_\mu u^\mu = \int \dd[4]{x} \sqrt{g} u_\mu \int \dd{s} u^\mu \frac{\delta^{(4)}[x-z(s)]}{\sqrt{g}},
\end{equation}
so the action \eqref{eq:pointaction_currents} can be written as
\begin{equation}
    S = -\int \dd[4]{x} \sqrt{g} \left( \frac{1}{4} F_{\mu\nu} F^{\mu\nu} + j^\mu A_\mu + j_m^\mu u_\mu \right), \label{totalDiff}
\end{equation}
with the mass current
\begin{equation}
    j_m^\mu = m \int \dd{s} u^\mu \frac{\delta^{(4)}[x-z(s)]}{\sqrt{g}}.
\end{equation}

From \eqref{totalDiff}, we can translate this action into a TDiff-invariant action for a charged massive particle subject to an electromagnetic field, just by replacing 
$\sqrt{g}$ by $f(g)$, namely
\begin{equation}
    S = -\int \dd[4]{x} f(g) \left( \frac{1}{4} F_{\mu\nu} F^{\mu\nu} + j^\mu A_\mu + j_m^\mu u_\mu \right).
\end{equation}

In order to obtain the Lorentz force for this TDiff-invariant setup,  we can consider the electromagnetic field to be external, ignoring possible backreaction, so the potential $A_\mu(x)$ is  fixed and we can ignore the electromagnetic part of the action. The rest of the action, after integrating the deltas out and changing the integration measure to coordinate time $\dd\tau$, can be written as
\begin{equation}
    S = -\int \dd{\tau} \frac{f(g)}{\sqrt{g}} \left( q v^\mu A_\mu + m \frac{\dd{s}}{\dd{\tau}}\right),
\end{equation}
where $v^\mu = \frac{\dd{z^\mu}}{\dd{\tau}} = (1, \vb{v})$ is the coordinate four-velocity, and $\frac{\dd{s}}{\dd{\tau}} = \sqrt{g_{\mu\nu}v^\mu v^\nu}$.

Variations of this action along the trajectory $\delta z^\alpha$ yield the following
\begin{align}
    \delta S = - \int \dd{\tau} \frac{f(g)}{\sqrt{g}} \delta z^\alpha &\left[ q v^\mu \partial_\alpha A_\mu - q \frac{\sqrt{g}}{f(g)}\frac{\dd}{\dd\tau} \left( \frac{f(g)}{\sqrt{g}} A_\alpha \right) + \frac{m}{2} \frac{(\partial_\alpha g_{\mu\nu}) v^\mu v^\nu}{\sqrt{g_{\mu\nu} v^\mu v^\nu}} \right. \nonumber \\ 
    &\left. - m \frac{\sqrt{g}}{f(g)}\frac{\dd}{\dd\tau} \left( \frac{f(g)}{\sqrt{g}} \frac{g_{\alpha\beta} v^\beta}{\sqrt{g_{\mu\nu}v^\mu v^\nu}} \right) + \partial_\alpha\left(\frac{f(g)}{\sqrt{g}}\right) \left( v^\mu A_\mu + m \frac{\dd{s}}{\dd{\tau}}\right) \right].
\end{align}

At leading order in the adiabatic expansion, i.e. neglecting terms involving metric 
derivatives, this yields the equation of motion
\begin{equation}
    m\, g_{\alpha\mu} \frac{\dd u^\mu}{\dd{\tau}} = q \left(v^\mu \partial_\alpha A_\mu - \frac{\dd{A_\alpha}}{\dd{\tau}} \right),
\end{equation}
which, as we could expect, does not depend on the TDiff function $f(g)$. The four-velocity, at the leading adiabatic order, can be written in a way that reminds of the Lorentz factor
\begin{equation}
    u^\mu = \frac{\dd z^\mu}{\dd{s}} = \frac{\dd z^\mu / \dd{\tau}}{\dd{s}/\dd{\tau}} = \frac{v^\mu}{\sqrt{1-\vb{v}^2}},
\end{equation}
while the total derivative of the vector field can be expanded as
\begin{equation}
    \frac{\dd{A_\alpha}}{\dd{\tau}} = \frac{\partial A_\alpha}{\partial \tau} + v^j \frac{\partial A_\alpha}{\partial x^j} = v^\mu \partial_\mu A_\alpha,
\end{equation}
which allows us to write the equation of motion in a compact manner
\begin{equation}
    m \frac{\dd u^\alpha}{\dd s} = q u_\mu F^{\alpha\mu}.
\end{equation}
which is nothing but the Lorentz-Dirac equation.

Setting $\alpha=i$, one finds the following equation of motion for the RW background
\begin{equation}
    \frac{\dd}{\dd\tau} \frac{m v^i}{\sqrt{1-\vb{v}^2}} = \frac{q}{a^2} ( \partial_0 A_i - \partial_i A_0 + v^j \partial_j A_i - v^j \partial_i A_j).
\end{equation}
After applying some vector identities and identifying the electric and magnetic fields \eqref{eq:def_EB}, this equation can be rewritten as
\begin{equation}
    \frac{\dd}{\dd\tau} \frac{m \vb{v}}{\sqrt{1-\vb{v}^2}} = \frac{q}{a^2} (\vb{E} + \vb{v}\cross \vb{B}),
\end{equation}
which is the Lorentz force law for a relativistic particle. On the other hand, if we set $\alpha=0$, we can obtain the equation for the variation of energy
\begin{equation}
    \frac{\dd}{\dd\tau} \frac{m}{\sqrt{1-\vb{v}^2}} = \frac{q}{b^2} \vb{v}\cdot\vb{E}.
\end{equation}

\bibliographystyle{apsrev4-1}
\bibliography{tdiff-vec.bib}

%merlin.mbs apsrev4-1.bst 2010-07-25 4.21a (PWD, AO, DPC) hacked
%Control: key (0)
%Control: author (72) initials jnrlst
%Control: editor formatted (1) identically to author
%Control: production of article title (-1) disabled
%Control: page (0) single
%Control: year (1) truncated
%Control: production of eprint (0) enabled
\begin{thebibliography}{46}%
\makeatletter
\providecommand \@ifxundefined [1]{%
 \@ifx{#1\undefined}
}%
\providecommand \@ifnum [1]{%
 \ifnum #1\expandafter \@firstoftwo
 \else \expandafter \@secondoftwo
 \fi
}%
\providecommand \@ifx [1]{%
 \ifx #1\expandafter \@firstoftwo
 \else \expandafter \@secondoftwo
 \fi
}%
\providecommand \natexlab [1]{#1}%
\providecommand \enquote  [1]{``#1''}%
\providecommand \bibnamefont  [1]{#1}%
\providecommand \bibfnamefont [1]{#1}%
\providecommand \citenamefont [1]{#1}%
\providecommand \href@noop [0]{\@secondoftwo}%
\providecommand \href [0]{\begingroup \@sanitize@url \@href}%
\providecommand \@href[1]{\@@startlink{#1}\@@href}%
\providecommand \@@href[1]{\endgroup#1\@@endlink}%
\providecommand \@sanitize@url [0]{\catcode `\\12\catcode `\$12\catcode `\&12\catcode `\#12\catcode `\^12\catcode `\_12\catcode `\%12\relax}%
\providecommand \@@startlink[1]{}%
\providecommand \@@endlink[0]{}%
\providecommand \url  [0]{\begingroup\@sanitize@url \@url }%
\providecommand \@url [1]{\endgroup\@href {#1}{\urlprefix }}%
\providecommand \urlprefix  [0]{URL }%
\providecommand \Eprint [0]{\href }%
\providecommand \doibase [0]{http://dx.doi.org/}%
\providecommand \selectlanguage [0]{\@gobble}%
\providecommand \bibinfo  [0]{\@secondoftwo}%
\providecommand \bibfield  [0]{\@secondoftwo}%
\providecommand \translation [1]{[#1]}%
\providecommand \BibitemOpen [0]{}%
\providecommand \bibitemStop [0]{}%
\providecommand \bibitemNoStop [0]{.\EOS\space}%
\providecommand \EOS [0]{\spacefactor3000\relax}%
\providecommand \BibitemShut  [1]{\csname bibitem#1\endcsname}%
\let\auto@bib@innerbib\@empty
%</preamble>
\bibitem [{\citenamefont {Will}(2014)}]{Will:2014kxa}%
  \BibitemOpen
  \bibfield  {author} {\bibinfo {author} {\bibfnamefont {C.~M.}\ \bibnamefont {Will}},\ }\href {\doibase 10.12942/lrr-2014-4} {\bibfield  {journal} {\bibinfo  {journal} {Living Rev. Rel.}\ }\textbf {\bibinfo {volume} {17}},\ \bibinfo {pages} {4} (\bibinfo {year} {2014})},\ \Eprint {http://arxiv.org/abs/1403.7377} {arXiv:1403.7377 [gr-qc]} \BibitemShut {NoStop}%
\bibitem [{\citenamefont {Alvarez}\ \emph {et~al.}(2006)\citenamefont {Alvarez}, \citenamefont {Blas}, \citenamefont {Garriga},\ and\ \citenamefont {Verdaguer}}]{Alvarez:2006uu}%
  \BibitemOpen
  \bibfield  {author} {\bibinfo {author} {\bibfnamefont {E.}~\bibnamefont {Alvarez}}, \bibinfo {author} {\bibfnamefont {D.}~\bibnamefont {Blas}}, \bibinfo {author} {\bibfnamefont {J.}~\bibnamefont {Garriga}}, \ and\ \bibinfo {author} {\bibfnamefont {E.}~\bibnamefont {Verdaguer}},\ }\href {\doibase 10.1016/j.nuclphysb.2006.08.003} {\bibfield  {journal} {\bibinfo  {journal} {Nucl. Phys. B}\ }\textbf {\bibinfo {volume} {756}},\ \bibinfo {pages} {148} (\bibinfo {year} {2006})},\ \Eprint {http://arxiv.org/abs/hep-th/0606019} {arXiv:hep-th/0606019} \BibitemShut {NoStop}%
\bibitem [{\citenamefont {Alvarez}\ \emph {et~al.}(2008)\citenamefont {Alvarez}, \citenamefont {Faedo},\ and\ \citenamefont {Lopez-Villarejo}}]{Alvarez:2008zw}%
  \BibitemOpen
  \bibfield  {author} {\bibinfo {author} {\bibfnamefont {E.}~\bibnamefont {Alvarez}}, \bibinfo {author} {\bibfnamefont {A.~F.}\ \bibnamefont {Faedo}}, \ and\ \bibinfo {author} {\bibfnamefont {J.~J.}\ \bibnamefont {Lopez-Villarejo}},\ }\href {\doibase 10.1088/1126-6708/2008/10/023} {\bibfield  {journal} {\bibinfo  {journal} {JHEP}\ }\textbf {\bibinfo {volume} {10}},\ \bibinfo {pages} {023} (\bibinfo {year} {2008})},\ \Eprint {http://arxiv.org/abs/0807.1293} {arXiv:0807.1293 [hep-th]} \BibitemShut {NoStop}%
\bibitem [{\citenamefont {Einstein}(1919)}]{Einstein:1919gv}%
  \BibitemOpen
  \bibfield  {author} {\bibinfo {author} {\bibfnamefont {A.}~\bibnamefont {Einstein}},\ }\href@noop {} {\bibfield  {journal} {\bibinfo  {journal} {Sitzungsber. Preuss. Akad. Wiss. Berlin (Math. Phys. )}\ }\textbf {\bibinfo {volume} {1919}},\ \bibinfo {pages} {349} (\bibinfo {year} {1919})}\BibitemShut {NoStop}%
\bibitem [{\citenamefont {Unruh}(1989)}]{Unruh:1988in}%
  \BibitemOpen
  \bibfield  {author} {\bibinfo {author} {\bibfnamefont {W.~G.}\ \bibnamefont {Unruh}},\ }\href {\doibase 10.1103/PhysRevD.40.1048} {\bibfield  {journal} {\bibinfo  {journal} {Phys. Rev. D}\ }\textbf {\bibinfo {volume} {40}},\ \bibinfo {pages} {1048} (\bibinfo {year} {1989})}\BibitemShut {NoStop}%
\bibitem [{\citenamefont {Henneaux}\ and\ \citenamefont {Teitelboim}(1989)}]{Henneaux:1989zc}%
  \BibitemOpen
  \bibfield  {author} {\bibinfo {author} {\bibfnamefont {M.}~\bibnamefont {Henneaux}}\ and\ \bibinfo {author} {\bibfnamefont {C.}~\bibnamefont {Teitelboim}},\ }\href {\doibase 10.1016/0370-2693(89)91251-3} {\bibfield  {journal} {\bibinfo  {journal} {Phys. Lett. B}\ }\textbf {\bibinfo {volume} {222}},\ \bibinfo {pages} {195} (\bibinfo {year} {1989})}\BibitemShut {NoStop}%
\bibitem [{\citenamefont {Kuchar}(1991)}]{Kuchar:1991xd}%
  \BibitemOpen
  \bibfield  {author} {\bibinfo {author} {\bibfnamefont {K.~V.}\ \bibnamefont {Kuchar}},\ }\href {\doibase 10.1103/PhysRevD.43.3332} {\bibfield  {journal} {\bibinfo  {journal} {Phys. Rev. D}\ }\textbf {\bibinfo {volume} {43}},\ \bibinfo {pages} {3332} (\bibinfo {year} {1991})}\BibitemShut {NoStop}%
\bibitem [{\citenamefont {Alvarez}(2005)}]{Alvarez:2005iy}%
  \BibitemOpen
  \bibfield  {author} {\bibinfo {author} {\bibfnamefont {E.}~\bibnamefont {Alvarez}},\ }\href {\doibase 10.1088/1126-6708/2005/03/002} {\bibfield  {journal} {\bibinfo  {journal} {JHEP}\ }\textbf {\bibinfo {volume} {03}},\ \bibinfo {pages} {002} (\bibinfo {year} {2005})},\ \Eprint {http://arxiv.org/abs/hep-th/0501146} {arXiv:hep-th/0501146} \BibitemShut {NoStop}%
\bibitem [{\citenamefont {Jirou\v{s}ek}\ and\ \citenamefont {Vikman}(2019)}]{Jirousek:2018ago}%
  \BibitemOpen
  \bibfield  {author} {\bibinfo {author} {\bibfnamefont {P.}~\bibnamefont {Jirou\v{s}ek}}\ and\ \bibinfo {author} {\bibfnamefont {A.}~\bibnamefont {Vikman}},\ }\href {\doibase 10.1088/1475-7516/2019/04/004} {\bibfield  {journal} {\bibinfo  {journal} {JCAP}\ }\textbf {\bibinfo {volume} {04}},\ \bibinfo {pages} {004} (\bibinfo {year} {2019})},\ \Eprint {http://arxiv.org/abs/1811.09547} {arXiv:1811.09547 [gr-qc]} \BibitemShut {NoStop}%
\bibitem [{\citenamefont {Hammer}\ \emph {et~al.}(2020)\citenamefont {Hammer}, \citenamefont {Jirousek},\ and\ \citenamefont {Vikman}}]{Hammer:2020dqp}%
  \BibitemOpen
  \bibfield  {author} {\bibinfo {author} {\bibfnamefont {K.}~\bibnamefont {Hammer}}, \bibinfo {author} {\bibfnamefont {P.}~\bibnamefont {Jirousek}}, \ and\ \bibinfo {author} {\bibfnamefont {A.}~\bibnamefont {Vikman}},\ }\href@noop {} {\  (\bibinfo {year} {2020})},\ \Eprint {http://arxiv.org/abs/2001.03169} {arXiv:2001.03169 [gr-qc]} \BibitemShut {NoStop}%
\bibitem [{\citenamefont {Carballo-Rubio}\ \emph {et~al.}(2022)\citenamefont {Carballo-Rubio}, \citenamefont {Garay},\ and\ \citenamefont {Garc\'\i{}a-Moreno}}]{Carballo-Rubio:2022ofy}%
  \BibitemOpen
  \bibfield  {author} {\bibinfo {author} {\bibfnamefont {R.}~\bibnamefont {Carballo-Rubio}}, \bibinfo {author} {\bibfnamefont {L.~J.}\ \bibnamefont {Garay}}, \ and\ \bibinfo {author} {\bibfnamefont {G.}~\bibnamefont {Garc\'\i{}a-Moreno}},\ }\href {\doibase 10.1088/1361-6382/aca386} {\bibfield  {journal} {\bibinfo  {journal} {Class. Quant. Grav.}\ }\textbf {\bibinfo {volume} {39}},\ \bibinfo {pages} {243001} (\bibinfo {year} {2022})},\ \Eprint {http://arxiv.org/abs/2207.08499} {arXiv:2207.08499 [gr-qc]} \BibitemShut {NoStop}%
\bibitem [{\citenamefont {Ellis}\ \emph {et~al.}(2011)\citenamefont {Ellis}, \citenamefont {van Elst}, \citenamefont {Murugan},\ and\ \citenamefont {Uzan}}]{Ellis:2010uc}%
  \BibitemOpen
  \bibfield  {author} {\bibinfo {author} {\bibfnamefont {G.~F.~R.}\ \bibnamefont {Ellis}}, \bibinfo {author} {\bibfnamefont {H.}~\bibnamefont {van Elst}}, \bibinfo {author} {\bibfnamefont {J.}~\bibnamefont {Murugan}}, \ and\ \bibinfo {author} {\bibfnamefont {J.-P.}\ \bibnamefont {Uzan}},\ }\href {\doibase 10.1088/0264-9381/28/22/225007} {\bibfield  {journal} {\bibinfo  {journal} {Class. Quant. Grav.}\ }\textbf {\bibinfo {volume} {28}},\ \bibinfo {pages} {225007} (\bibinfo {year} {2011})},\ \Eprint {http://arxiv.org/abs/1008.1196} {arXiv:1008.1196 [gr-qc]} \BibitemShut {NoStop}%
\bibitem [{\citenamefont {Pirogov}(2006)}]{Pirogov:2005im}%
  \BibitemOpen
  \bibfield  {author} {\bibinfo {author} {\bibfnamefont {Y.~F.}\ \bibnamefont {Pirogov}},\ }\href {\doibase 10.1134/S1063778806080102} {\bibfield  {journal} {\bibinfo  {journal} {Phys. Atom. Nucl.}\ }\textbf {\bibinfo {volume} {69}},\ \bibinfo {pages} {1338} (\bibinfo {year} {2006})},\ \Eprint {http://arxiv.org/abs/gr-qc/0505031} {arXiv:gr-qc/0505031} \BibitemShut {NoStop}%
\bibitem [{\citenamefont {Pirogov}(2010)}]{Pirogov:2009hr}%
  \BibitemOpen
  \bibfield  {author} {\bibinfo {author} {\bibfnamefont {Y.~F.}\ \bibnamefont {Pirogov}},\ }\href {\doibase 10.1134/S1063778810010151} {\bibfield  {journal} {\bibinfo  {journal} {Phys. Atom. Nucl.}\ }\textbf {\bibinfo {volume} {73}},\ \bibinfo {pages} {134} (\bibinfo {year} {2010})},\ \Eprint {http://arxiv.org/abs/0903.2018} {arXiv:0903.2018 [gr-qc]} \BibitemShut {NoStop}%
\bibitem [{\citenamefont {Pirogov}(2012)}]{Pirogov:2011iq}%
  \BibitemOpen
  \bibfield  {author} {\bibinfo {author} {\bibfnamefont {Y.~F.}\ \bibnamefont {Pirogov}},\ }\href {\doibase 10.1140/epjc/s10052-012-2017-y} {\bibfield  {journal} {\bibinfo  {journal} {Eur. Phys. J. C}\ }\textbf {\bibinfo {volume} {72}},\ \bibinfo {pages} {2017} (\bibinfo {year} {2012})},\ \Eprint {http://arxiv.org/abs/1111.1437} {arXiv:1111.1437 [gr-qc]} \BibitemShut {NoStop}%
\bibitem [{\citenamefont {Pirogov}(2015)}]{Pirogov:2014lda}%
  \BibitemOpen
  \bibfield  {author} {\bibinfo {author} {\bibfnamefont {Y.~F.}\ \bibnamefont {Pirogov}},\ }\href {\doibase 10.1134/S1063778815030084} {\bibfield  {journal} {\bibinfo  {journal} {Phys. Atom. Nucl.}\ }\textbf {\bibinfo {volume} {78}},\ \bibinfo {pages} {528} (\bibinfo {year} {2015})},\ \Eprint {http://arxiv.org/abs/1401.8191} {arXiv:1401.8191 [gr-qc]} \BibitemShut {NoStop}%
\bibitem [{\citenamefont {Bello-Morales}\ and\ \citenamefont {Maroto}(2024)}]{Bello-Morales:2023btf}%
  \BibitemOpen
  \bibfield  {author} {\bibinfo {author} {\bibfnamefont {A.~G.}\ \bibnamefont {Bello-Morales}}\ and\ \bibinfo {author} {\bibfnamefont {A.~L.}\ \bibnamefont {Maroto}},\ }\href {\doibase 10.1103/PhysRevD.109.043506} {\bibfield  {journal} {\bibinfo  {journal} {Phys. Rev. D}\ }\textbf {\bibinfo {volume} {109}},\ \bibinfo {pages} {043506} (\bibinfo {year} {2024})},\ \Eprint {http://arxiv.org/abs/2308.00635} {arXiv:2308.00635 [gr-qc]} \BibitemShut {NoStop}%
\bibitem [{\citenamefont {Alvarez}\ and\ \citenamefont {Faedo}(2007)}]{Alvarez:2007nn}%
  \BibitemOpen
  \bibfield  {author} {\bibinfo {author} {\bibfnamefont {E.}~\bibnamefont {Alvarez}}\ and\ \bibinfo {author} {\bibfnamefont {A.~F.}\ \bibnamefont {Faedo}},\ }\href {\doibase 10.1103/PhysRevD.76.064013} {\bibfield  {journal} {\bibinfo  {journal} {Phys. Rev. D}\ }\textbf {\bibinfo {volume} {76}},\ \bibinfo {pages} {064013} (\bibinfo {year} {2007})},\ \Eprint {http://arxiv.org/abs/hep-th/0702184} {arXiv:hep-th/0702184} \BibitemShut {NoStop}%
\bibitem [{\citenamefont {Alvarez}\ \emph {et~al.}(2009)\citenamefont {Alvarez}, \citenamefont {Faedo},\ and\ \citenamefont {Lopez-Villarejo}}]{Alvarez:2009ga}%
  \BibitemOpen
  \bibfield  {author} {\bibinfo {author} {\bibfnamefont {E.}~\bibnamefont {Alvarez}}, \bibinfo {author} {\bibfnamefont {A.~F.}\ \bibnamefont {Faedo}}, \ and\ \bibinfo {author} {\bibfnamefont {J.~J.}\ \bibnamefont {Lopez-Villarejo}},\ }\href {\doibase 10.1088/1475-7516/2009/07/002} {\bibfield  {journal} {\bibinfo  {journal} {JCAP}\ }\textbf {\bibinfo {volume} {07}},\ \bibinfo {pages} {002} (\bibinfo {year} {2009})},\ \Eprint {http://arxiv.org/abs/0904.3298} {arXiv:0904.3298 [hep-th]} \BibitemShut {NoStop}%
\bibitem [{\citenamefont {Maroto}(2023)}]{Maroto:2023toq}%
  \BibitemOpen
  \bibfield  {author} {\bibinfo {author} {\bibfnamefont {A.~L.}\ \bibnamefont {Maroto}},\ }\href@noop {} {\  (\bibinfo {year} {2023})},\ \Eprint {http://arxiv.org/abs/2301.05713} {arXiv:2301.05713 [gr-qc]} \BibitemShut {NoStop}%
\bibitem [{\citenamefont {Alonso-L\'opez}\ \emph {et~al.}(2024)\citenamefont {Alonso-L\'opez}, \citenamefont {de~Cruz~P\'erez},\ and\ \citenamefont {Maroto}}]{Alonso-Lopez:2023hkx}%
  \BibitemOpen
  \bibfield  {author} {\bibinfo {author} {\bibfnamefont {D.}~\bibnamefont {Alonso-L\'opez}}, \bibinfo {author} {\bibfnamefont {J.}~\bibnamefont {de~Cruz~P\'erez}}, \ and\ \bibinfo {author} {\bibfnamefont {A.~L.}\ \bibnamefont {Maroto}},\ }\href {\doibase 10.1103/PhysRevD.109.023537} {\bibfield  {journal} {\bibinfo  {journal} {Phys. Rev. D}\ }\textbf {\bibinfo {volume} {109}},\ \bibinfo {pages} {023537} (\bibinfo {year} {2024})},\ \Eprint {http://arxiv.org/abs/2311.16836} {arXiv:2311.16836 [astro-ph.CO]} \BibitemShut {NoStop}%
\bibitem [{\citenamefont {Jaramillo-Garrido}\ \emph {et~al.}(2023)\citenamefont {Jaramillo-Garrido}, \citenamefont {Maroto},\ and\ \citenamefont {Mart\'\i{}n-Moruno}}]{Jaramillo-Garrido:2023cor}%
  \BibitemOpen
  \bibfield  {author} {\bibinfo {author} {\bibfnamefont {D.}~\bibnamefont {Jaramillo-Garrido}}, \bibinfo {author} {\bibfnamefont {A.~L.}\ \bibnamefont {Maroto}}, \ and\ \bibinfo {author} {\bibfnamefont {P.}~\bibnamefont {Mart\'\i{}n-Moruno}},\ }\href@noop {} {\  (\bibinfo {year} {2023})},\ \Eprint {http://arxiv.org/abs/2307.14861} {arXiv:2307.14861 [gr-qc]} \BibitemShut {NoStop}%
\bibitem [{\citenamefont {Jaramillo-Garrido}\ \emph {et~al.}(2024)\citenamefont {Jaramillo-Garrido}, \citenamefont {Maroto},\ and\ \citenamefont {Mart\'\i{}n-Moruno}}]{Jaramillo-Garrido:2024tdv}%
  \BibitemOpen
  \bibfield  {author} {\bibinfo {author} {\bibfnamefont {D.}~\bibnamefont {Jaramillo-Garrido}}, \bibinfo {author} {\bibfnamefont {A.~L.}\ \bibnamefont {Maroto}}, \ and\ \bibinfo {author} {\bibfnamefont {P.}~\bibnamefont {Mart\'\i{}n-Moruno}},\ }\href@noop {} {\  (\bibinfo {year} {2024})},\ \Eprint {http://arxiv.org/abs/2402.17422} {arXiv:2402.17422 [gr-qc]} \BibitemShut {NoStop}%
\bibitem [{\citenamefont {Koivisto}\ and\ \citenamefont {Mota}(2008)}]{Koivisto:2008xf}%
  \BibitemOpen
  \bibfield  {author} {\bibinfo {author} {\bibfnamefont {T.}~\bibnamefont {Koivisto}}\ and\ \bibinfo {author} {\bibfnamefont {D.~F.}\ \bibnamefont {Mota}},\ }\href {\doibase 10.1088/1475-7516/2008/08/021} {\bibfield  {journal} {\bibinfo  {journal} {JCAP}\ }\textbf {\bibinfo {volume} {08}},\ \bibinfo {pages} {021} (\bibinfo {year} {2008})},\ \Eprint {http://arxiv.org/abs/0805.4229} {arXiv:0805.4229 [astro-ph]} \BibitemShut {NoStop}%
\bibitem [{\citenamefont {Maleknejad}\ and\ \citenamefont {Sheikh-Jabbari}(2013)}]{Maleknejad:2011jw}%
  \BibitemOpen
  \bibfield  {author} {\bibinfo {author} {\bibfnamefont {A.}~\bibnamefont {Maleknejad}}\ and\ \bibinfo {author} {\bibfnamefont {M.~M.}\ \bibnamefont {Sheikh-Jabbari}},\ }\href {\doibase 10.1016/j.physletb.2013.05.001} {\bibfield  {journal} {\bibinfo  {journal} {Phys. Lett. B}\ }\textbf {\bibinfo {volume} {723}},\ \bibinfo {pages} {224} (\bibinfo {year} {2013})},\ \Eprint {http://arxiv.org/abs/1102.1513} {arXiv:1102.1513 [hep-ph]} \BibitemShut {NoStop}%
\bibitem [{\citenamefont {Maleknejad}\ \emph {et~al.}(2013)\citenamefont {Maleknejad}, \citenamefont {Sheikh-Jabbari},\ and\ \citenamefont {Soda}}]{Maleknejad:2012fw}%
  \BibitemOpen
  \bibfield  {author} {\bibinfo {author} {\bibfnamefont {A.}~\bibnamefont {Maleknejad}}, \bibinfo {author} {\bibfnamefont {M.~M.}\ \bibnamefont {Sheikh-Jabbari}}, \ and\ \bibinfo {author} {\bibfnamefont {J.}~\bibnamefont {Soda}},\ }\href {\doibase 10.1016/j.physrep.2013.03.003} {\bibfield  {journal} {\bibinfo  {journal} {Phys. Rept.}\ }\textbf {\bibinfo {volume} {528}},\ \bibinfo {pages} {161} (\bibinfo {year} {2013})},\ \Eprint {http://arxiv.org/abs/1212.2921} {arXiv:1212.2921 [hep-th]} \BibitemShut {NoStop}%
\bibitem [{\citenamefont {Armendariz-Picon}(2004)}]{Armendariz-Picon:2004say}%
  \BibitemOpen
  \bibfield  {author} {\bibinfo {author} {\bibfnamefont {C.}~\bibnamefont {Armendariz-Picon}},\ }\href {\doibase 10.1088/1475-7516/2004/07/007} {\bibfield  {journal} {\bibinfo  {journal} {JCAP}\ }\textbf {\bibinfo {volume} {07}},\ \bibinfo {pages} {007} (\bibinfo {year} {2004})},\ \Eprint {http://arxiv.org/abs/astro-ph/0405267} {arXiv:astro-ph/0405267} \BibitemShut {NoStop}%
\bibitem [{\citenamefont {Boehmer}\ and\ \citenamefont {Harko}(2007)}]{Boehmer:2007qa}%
  \BibitemOpen
  \bibfield  {author} {\bibinfo {author} {\bibfnamefont {C.~G.}\ \bibnamefont {Boehmer}}\ and\ \bibinfo {author} {\bibfnamefont {T.}~\bibnamefont {Harko}},\ }\href {\doibase 10.1140/epjc/s10052-007-0210-1} {\bibfield  {journal} {\bibinfo  {journal} {Eur. Phys. J. C}\ }\textbf {\bibinfo {volume} {50}},\ \bibinfo {pages} {423} (\bibinfo {year} {2007})},\ \Eprint {http://arxiv.org/abs/gr-qc/0701029} {arXiv:gr-qc/0701029} \BibitemShut {NoStop}%
\bibitem [{\citenamefont {Beltran~Jimenez}\ and\ \citenamefont {Maroto}(2008)}]{BeltranJimenez:2008iye}%
  \BibitemOpen
  \bibfield  {author} {\bibinfo {author} {\bibfnamefont {J.}~\bibnamefont {Beltran~Jimenez}}\ and\ \bibinfo {author} {\bibfnamefont {A.~L.}\ \bibnamefont {Maroto}},\ }\href {\doibase 10.1103/PhysRevD.78.063005} {\bibfield  {journal} {\bibinfo  {journal} {Phys. Rev. D}\ }\textbf {\bibinfo {volume} {78}},\ \bibinfo {pages} {063005} (\bibinfo {year} {2008})},\ \Eprint {http://arxiv.org/abs/0801.1486} {arXiv:0801.1486 [astro-ph]} \BibitemShut {NoStop}%
\bibitem [{\citenamefont {Beltran~Jimenez}\ and\ \citenamefont {Maroto}(2009)}]{BeltranJimenez:2008enx}%
  \BibitemOpen
  \bibfield  {author} {\bibinfo {author} {\bibfnamefont {J.}~\bibnamefont {Beltran~Jimenez}}\ and\ \bibinfo {author} {\bibfnamefont {A.~L.}\ \bibnamefont {Maroto}},\ }\href {\doibase 10.1088/1475-7516/2009/03/016} {\bibfield  {journal} {\bibinfo  {journal} {JCAP}\ }\textbf {\bibinfo {volume} {03}},\ \bibinfo {pages} {016} (\bibinfo {year} {2009})},\ \Eprint {http://arxiv.org/abs/0811.0566} {arXiv:0811.0566 [astro-ph]} \BibitemShut {NoStop}%
\bibitem [{\citenamefont {Nelson}\ and\ \citenamefont {Scholtz}(2011)}]{Nelson:2011sf}%
  \BibitemOpen
  \bibfield  {author} {\bibinfo {author} {\bibfnamefont {A.~E.}\ \bibnamefont {Nelson}}\ and\ \bibinfo {author} {\bibfnamefont {J.}~\bibnamefont {Scholtz}},\ }\href {\doibase 10.1103/PhysRevD.84.103501} {\bibfield  {journal} {\bibinfo  {journal} {Phys. Rev. D}\ }\textbf {\bibinfo {volume} {84}},\ \bibinfo {pages} {103501} (\bibinfo {year} {2011})},\ \Eprint {http://arxiv.org/abs/1105.2812} {arXiv:1105.2812 [hep-ph]} \BibitemShut {NoStop}%
\bibitem [{\citenamefont {{L{\'{o}}pez Nacir}}\ and\ \citenamefont {Urban}(2018)}]{LopezNacir:2018epg}%
  \BibitemOpen
  \bibfield  {author} {\bibinfo {author} {\bibfnamefont {D.}~\bibnamefont {{L{\'{o}}pez Nacir}}}\ and\ \bibinfo {author} {\bibfnamefont {F.~R.}\ \bibnamefont {Urban}},\ }\href {\doibase 10.1088/1475-7516/2018/10/044} {\bibfield  {journal} {\bibinfo  {journal} {JCAP}\ }\textbf {\bibinfo {volume} {10}},\ \bibinfo {pages} {44} (\bibinfo {year} {2018})},\ \Eprint {http://arxiv.org/abs/1807.10491} {arXiv:1807.10491 [astro-ph.CO]} \BibitemShut {NoStop}%
\bibitem [{\citenamefont {Cembranos}\ \emph {et~al.}(2017)\citenamefont {Cembranos}, \citenamefont {Maroto},\ and\ \citenamefont {{N{\'{u}}{\~{n}}ez Jare{\~{n}}o}}}]{Cembranos:2016ugq}%
  \BibitemOpen
  \bibfield  {author} {\bibinfo {author} {\bibfnamefont {J.~A.~R.}\ \bibnamefont {Cembranos}}, \bibinfo {author} {\bibfnamefont {A.~L.}\ \bibnamefont {Maroto}}, \ and\ \bibinfo {author} {\bibfnamefont {S.~J.}\ \bibnamefont {{N{\'{u}}{\~{n}}ez Jare{\~{n}}o}}},\ }\href {\doibase 10.1007/JHEP02(2017)064} {\bibfield  {journal} {\bibinfo  {journal} {JHEP}\ }\textbf {\bibinfo {volume} {02}},\ \bibinfo {pages} {64} (\bibinfo {year} {2017})},\ \Eprint {http://arxiv.org/abs/1611.03793} {arXiv:1611.03793 [astro-ph.CO]} \BibitemShut {NoStop}%
\bibitem [{\citenamefont {Blas}\ \emph {et~al.}(2011)\citenamefont {Blas}, \citenamefont {Shaposhnikov},\ and\ \citenamefont {Zenhausern}}]{Blas:2011ac}%
  \BibitemOpen
  \bibfield  {author} {\bibinfo {author} {\bibfnamefont {D.}~\bibnamefont {Blas}}, \bibinfo {author} {\bibfnamefont {M.}~\bibnamefont {Shaposhnikov}}, \ and\ \bibinfo {author} {\bibfnamefont {D.}~\bibnamefont {Zenhausern}},\ }\href {\doibase 10.1103/PhysRevD.84.044001} {\bibfield  {journal} {\bibinfo  {journal} {Phys. Rev. D}\ }\textbf {\bibinfo {volume} {84}},\ \bibinfo {pages} {044001} (\bibinfo {year} {2011})},\ \Eprint {http://arxiv.org/abs/1104.1392} {arXiv:1104.1392 [hep-th]} \BibitemShut {NoStop}%
\bibitem [{\citenamefont {Rubakov}(2014)}]{Rubakov:2014jja}%
  \BibitemOpen
  \bibfield  {author} {\bibinfo {author} {\bibfnamefont {V.~A.}\ \bibnamefont {Rubakov}},\ }\href {\doibase 10.3367/UFNe.0184.201402b.0137} {\bibfield  {journal} {\bibinfo  {journal} {Phys. Usp.}\ }\textbf {\bibinfo {volume} {57}},\ \bibinfo {pages} {128} (\bibinfo {year} {2014})},\ \Eprint {http://arxiv.org/abs/1401.4024} {arXiv:1401.4024 [hep-th]} \BibitemShut {NoStop}%
\bibitem [{\citenamefont {Misner}\ \emph {et~al.}(1973)\citenamefont {Misner}, \citenamefont {Thorne},\ and\ \citenamefont {Wheeler}}]{Misner:1973prb}%
  \BibitemOpen
  \bibfield  {author} {\bibinfo {author} {\bibfnamefont {C.~W.}\ \bibnamefont {Misner}}, \bibinfo {author} {\bibfnamefont {K.~S.}\ \bibnamefont {Thorne}}, \ and\ \bibinfo {author} {\bibfnamefont {J.~A.}\ \bibnamefont {Wheeler}},\ }\href@noop {} {\emph {\bibinfo {title} {{Gravitation}}}}\ (\bibinfo  {publisher} {W. H. Freeman},\ \bibinfo {address} {San Francisco},\ \bibinfo {year} {1973})\BibitemShut {NoStop}%
\bibitem [{\citenamefont {Birrell}\ and\ \citenamefont {Davies}(1984)}]{Birrell:1982ix}%
  \BibitemOpen
  \bibfield  {author} {\bibinfo {author} {\bibfnamefont {N.~D.}\ \bibnamefont {Birrell}}\ and\ \bibinfo {author} {\bibfnamefont {P.~C.~W.}\ \bibnamefont {Davies}},\ }\href {\doibase 10.1017/CBO9780511622632} {\emph {\bibinfo {title} {{Quantum Fields in Curved Space}}}},\ Cambridge Monographs on Mathematical Physics\ (\bibinfo  {publisher} {Cambridge Univ. Press},\ \bibinfo {address} {Cambridge, UK},\ \bibinfo {year} {1984})\BibitemShut {NoStop}%
\bibitem [{\citenamefont {Nakahara}(2003)}]{Nakahara:2003nw}%
  \BibitemOpen
  \bibfield  {author} {\bibinfo {author} {\bibfnamefont {M.}~\bibnamefont {Nakahara}},\ }\href@noop {} {\emph {\bibinfo {title} {{Geometry, topology and physics}}}}\ (\bibinfo {year} {2003})\BibitemShut {NoStop}%
\bibitem [{\citenamefont {Shapiro}(2022)}]{Shapiro:2016pfm}%
  \BibitemOpen
  \bibfield  {author} {\bibinfo {author} {\bibfnamefont {I.~L.}\ \bibnamefont {Shapiro}},\ }\href {\doibase 10.3390/universe8110586} {\bibfield  {journal} {\bibinfo  {journal} {Universe}\ }\textbf {\bibinfo {volume} {8}},\ \bibinfo {pages} {586} (\bibinfo {year} {2022})},\ \Eprint {http://arxiv.org/abs/1611.02263} {arXiv:1611.02263 [gr-qc]} \BibitemShut {NoStop}%
\bibitem [{\citenamefont {Dobado}\ \emph {et~al.}(1997)\citenamefont {Dobado}, \citenamefont {Gomez-Nicola}, \citenamefont {Maroto},\ and\ \citenamefont {Pelaez}}]{Dobado:1997jx}%
  \BibitemOpen
  \bibfield  {author} {\bibinfo {author} {\bibfnamefont {A.}~\bibnamefont {Dobado}}, \bibinfo {author} {\bibfnamefont {A.}~\bibnamefont {Gomez-Nicola}}, \bibinfo {author} {\bibfnamefont {A.~L.}\ \bibnamefont {Maroto}}, \ and\ \bibinfo {author} {\bibfnamefont {J.~R.}\ \bibnamefont {Pelaez}},\ }\href@noop {} {\emph {\bibinfo {title} {{Effective lagrangians for the standard model}}}}\ (\bibinfo {year} {1997})\BibitemShut {NoStop}%
\bibitem [{\citenamefont {Oancea}\ and\ \citenamefont {Kumar}(2023)}]{Oancea:2022utx}%
  \BibitemOpen
  \bibfield  {author} {\bibinfo {author} {\bibfnamefont {M.~A.}\ \bibnamefont {Oancea}}\ and\ \bibinfo {author} {\bibfnamefont {A.}~\bibnamefont {Kumar}},\ }\href {\doibase 10.1103/PhysRevD.107.044029} {\bibfield  {journal} {\bibinfo  {journal} {Phys. Rev. D}\ }\textbf {\bibinfo {volume} {107}},\ \bibinfo {pages} {044029} (\bibinfo {year} {2023})},\ \Eprint {http://arxiv.org/abs/2212.04414} {arXiv:2212.04414 [gr-qc]} \BibitemShut {NoStop}%
\bibitem [{\citenamefont {Itzykson}\ and\ \citenamefont {Zuber}(1980)}]{Itzykson:1980rh}%
  \BibitemOpen
  \bibfield  {author} {\bibinfo {author} {\bibfnamefont {C.}~\bibnamefont {Itzykson}}\ and\ \bibinfo {author} {\bibfnamefont {J.~B.}\ \bibnamefont {Zuber}},\ }\href@noop {} {\emph {\bibinfo {title} {{Quantum Field Theory}}}},\ International Series In Pure and Applied Physics\ (\bibinfo  {publisher} {McGraw-Hill},\ \bibinfo {address} {New York},\ \bibinfo {year} {1980})\BibitemShut {NoStop}%
\bibitem [{\citenamefont {Pfenning}(2002)}]{Pfenning:2001wx}%
  \BibitemOpen
  \bibfield  {author} {\bibinfo {author} {\bibfnamefont {M.~J.}\ \bibnamefont {Pfenning}},\ }\href {\doibase 10.1103/PhysRevD.65.024009} {\bibfield  {journal} {\bibinfo  {journal} {Phys. Rev. D}\ }\textbf {\bibinfo {volume} {65}},\ \bibinfo {pages} {024009} (\bibinfo {year} {2002})},\ \Eprint {http://arxiv.org/abs/gr-qc/0107075} {arXiv:gr-qc/0107075} \BibitemShut {NoStop}%
\bibitem [{\citenamefont {Beltran~Jimenez}\ and\ \citenamefont {Maroto}(2010)}]{BeltranJimenez:2009oao}%
  \BibitemOpen
  \bibfield  {author} {\bibinfo {author} {\bibfnamefont {J.}~\bibnamefont {Beltran~Jimenez}}\ and\ \bibinfo {author} {\bibfnamefont {A.~L.}\ \bibnamefont {Maroto}},\ }\href {\doibase 10.1016/j.physletb.2010.02.038} {\bibfield  {journal} {\bibinfo  {journal} {Phys. Lett. B}\ }\textbf {\bibinfo {volume} {686}},\ \bibinfo {pages} {175} (\bibinfo {year} {2010})},\ \Eprint {http://arxiv.org/abs/0903.4672} {arXiv:0903.4672 [astro-ph.CO]} \BibitemShut {NoStop}%
\bibitem [{\citenamefont {Turner}\ and\ \citenamefont {Widrow}(1988)}]{Turner:1987bw}%
  \BibitemOpen
  \bibfield  {author} {\bibinfo {author} {\bibfnamefont {M.~S.}\ \bibnamefont {Turner}}\ and\ \bibinfo {author} {\bibfnamefont {L.~M.}\ \bibnamefont {Widrow}},\ }\href {\doibase 10.1103/PhysRevD.37.2743} {\bibfield  {journal} {\bibinfo  {journal} {Phys. Rev. D}\ }\textbf {\bibinfo {volume} {37}},\ \bibinfo {pages} {2743} (\bibinfo {year} {1988})}\BibitemShut {NoStop}%
\bibitem [{\citenamefont {Durrer}\ and\ \citenamefont {Neronov}(2013)}]{Durrer:2013pga}%
  \BibitemOpen
  \bibfield  {author} {\bibinfo {author} {\bibfnamefont {R.}~\bibnamefont {Durrer}}\ and\ \bibinfo {author} {\bibfnamefont {A.}~\bibnamefont {Neronov}},\ }\href {\doibase 10.1007/s00159-013-0062-7} {\bibfield  {journal} {\bibinfo  {journal} {Astron. Astrophys. Rev.}\ }\textbf {\bibinfo {volume} {21}},\ \bibinfo {pages} {62} (\bibinfo {year} {2013})},\ \Eprint {http://arxiv.org/abs/1303.7121} {arXiv:1303.7121 [astro-ph.CO]} \BibitemShut {NoStop}%
\end{thebibliography}%

\end{document}